# The balance between paraelectricity and ferroelectricity in non-chiral smectic homologs

Dorota Węgłowska, Michał Czerwiński, Robert Dzienisiewicz, Paweł Perkowski, Jadwiga Szydłowska, Damian Pociecha, Mateusz Mrukiewicz*

Dorota Węgłowska, Michał Czerwiński, Robert Dzienisiewicz
*Institute of Chemistry, Military University of Technology, Kaliskiego 2, 00-908 Warsaw, Poland*

Paweł Perkowski, Mateusz Mrukiewicz
*Institute of Applied Physics, Military University of Technology, Kaliskiego 2, 00-908 Warsaw, Poland*

Damian Pociecha, Jadwiga Szydłowska
*Faculty of Chemistry, University of Warsaw, Żwirki i Wigury 101, 02-089 Warsaw, Poland*

Corresponding author: mateusz.mrukiewicz@wat.edu.pl

**Abstract**

Non-chiral liquid crystals (LCs) exhibiting ferroelectricity, distinguished by their dynamic responsiveness to external stimuli and high spontaneous polarization, provide renewed impetus for research into this area of soft matter and open novel application possibilities. Consequently, the identification of structural elements within LC compounds that promote ferroelectricity in non-chiral systems is of critical importance. In this work, two homologs of rod-like compounds, with phenyl and ester groups in the rigid core substituted by fluorine atoms, differing by a single methylene unit, were synthesized and comprehensively analyzed using complementary experimental techniques and quantum-mechanical modeling. This systematic study presents the first documented instance in which such a minimal structural modification markedly influences the polarity of smectic phases in two homologs without substantially altering the phase transition temperatures, particularly the sequence and temperature ranges of smectic and nematic phases. Furthermore, the findings reveal that the longer homolog, which exhibits paraelectric phases, demonstrates a pronounced capacity to maintain ferroelectric phases in mixtures. These results provide new insights into the critical structure-property relationships between molecular architecture and ferroelectric characteristics in LCs. Moreover, the properties of the studied mixtures underscore the potential to develop LC mixtures with ferroelectric properties in a broad temperature range, a feature of considerable significance for applications.

## Introduction

Ferroelectric liquid crystals (FLCs) in non-chiral systems have become the subject of intense and open scientific discussion in recent years due to their unique combination of physical properties, including spontaneous polarization, fast electro-optical switching, and non-linear optical properties.[1] These features make FLCs promising materials for advanced applications, especially in the area of display and imaging technologies. The recent discovery of new ferroelectric nematic LC phases has given rise to new trends and challenges related to utilizing their full potential in next-generation optical and electronic devices.[1] Moreover, there is still an open scientific question to be resolved regarding symmetry breaking and polar ordering in the non-chiral soft matter field. The emergence of a new ferroelectric phase in single compounds has allowed the discovery of new chiral and non-chiral polar phases.[2]

Polar materials have a permanent dipole moment which can order spontaneously to macroscopic polarization in the absence of external electric field *E*. This special property of materials is called ferroelectricity.[3] The resulting spontaneous polarization $P_S$ is sensitive to the sign of *E*. The confirmation of the occurrence of ferroelectricity in materials is the observation of a second-order nonlinear optical effect, which is evidence of the lack of centrosymmetry. In the beginning, ferroelectric properties were observed mainly in solid crystals.[4] However, in 1978, ferroelectricity was discovered in the chiral smectic C liquid crystal (SmC*),[5] which confirms Robert Meyer's theory about the consequences of the presence of chiral molecules in the layered mesophase.[6]

In the SmC phase, the director *n*, a local direction along the long molecular axes, is tilted by the angle of $\theta$ with respect to the smectic layer normal. Meyer theoretically predicted that incorporating chiral molecules into the SmC phase breaks the mirror symmetry within the structure. The chiral center in the rigid core of the molecule with a perpendicular dipole moment restricts the ability of the molecule to rotate around its long axis. As a result, in SmC* each individual layer possesses a local spontaneous polarization that rotates in a screw-like manner from layer to layer and a helical modulation of the director *n* appears. This polarization is rather low (tens or hundreds of $nC/cm^2$).[3] The liquid crystal compounds with SmC* usually also exhibit the higher temperature paraelectric phases: the chiral smectic A phase (SmA*) and the chiral nematic phase (N*). The layers in the non-chiral SmA phase are formed in such a way that the director is oriented along the normal to the layer. The non-chiral nematic N phase, compared to the smectic phase A, shows no translational order along *n*, but only long-range orientational order. The smectic C ordering may form an antiferroelectric structure called the $SmC_A$*.[7–9] In the anticlinic structure of the $SmC_A$* phase, the director tilt, and thus the sign of the spontaneous polarization, changes between adjacent layers.

Ferroelectricity in liquid crystals was initially thought to be related to molecular chirality. However, ferroelectric behavior was also observed in the non-chiral banana-shaped molecules.[10] The origin of ferroelectricity in this type of LCs is based on the $C_{2v}$ polar symmetry, which is obtained from efficient molecular packing. Due to the unique packing mechanism, the spontaneous polarization in such a system aligns parallel to the layer and is switchable by the applied electric field.

In 2017, the research groups independently and simultaneously reported ferroelectric properties in single compounds with non-chiral and rod-like structures, named DIO and RM734, which exhibit the uniaxial N phase,[11–13] namely $N_F$. This great discovery of the additional state of matter started intensive work on the synthesis of other novel liquid-crystalline polar materials. On the basis of the understanding of the molecular structure of these two ferroelectric compounds, many groundbreaking findings have been made in the field of new materials in this type of soft matter. Recently, it has been demonstrated that it is possible to obtain a thermodynamically stable $N_F$ phase,[14,15] which, in contrast to the phase occurring in the above-mentioned DIO and RM734, also appears upon heating above the melting temperature. Moreover, the phenomenon of the spontaneous mirror symmetry breaking, consisting of the formation of chiral structures from achiral molecules, was observed in the ferroelectric twist-bend nematic ($N_{TBF}$) phase.[16] In this structure, the molecules form a helicoid, where the polarization direction is consistent with the helicoidal axis. Another similar case is the structure of the polar heliconical smectic C ($SmC_P^H$) phase, where the polarization direction is oriented parallel to the layer normal.[17] Both $N_{TBF}$ and $SmC_P^H$ phases show the ability to observe the selective light reflection.

Further progress in $N_F$ phase research led to the discovery of the non-chiral ferroelectric smectic A ($SmA_F$) phase.[18–20] In $SmA_F$, the polarization direction, which runs along the director, is perpendicular to the planes of the smectic layer. In the layers, domains of opposite polarity are separated by polarization-reverse walls.[18] The emergence of the phase was observed in mixtures and as well in single compounds at transitions from the different polar and non-polar liquid crystalline phases.[2] The evidence of the ferroelectric smectic C phase ($SmC_F$) was observed in the temperature region below $SmA_F$.[21,22] In contrast to the well-known ferroelectric SmC*, the $SmC_F$ phase does not exhibit the helical order. Thus, the polarization vector is parallel to the director. Therefore, $SmC_F$ is considered to be a proper ferroelectric layered fluid because the polarization is due to the spontaneous polar orientation of the molecular dipoles and not to the hindered molecular rotation as in the case of SmC*. In the binary mixture of highly polar mesogenic materials, based on DIO, the antiferroelectric smectic A phase ($SmA_{AF}$) was observed.[23] Moreover, it has been demonstrated that the polar ordering of SmC can be found in compounds with significantly smaller dipole moments than typical ferroelectric fluids.[24] This suggests that a large dipole moment is not a determinant of the formation of ferroelectricity in non-chiral smectic liquid crystals.

Here, we present experimental results with quantum modeling of two newly synthesized non-chiral rod-like homologs, named 3F and 4F. These compounds, which have phenyl and ester groups with fluorine substitutions in rigid cores, differ only in one methylene unit in the terminal alkyl chain. This systematic study provides the first evidence that such a minimal structural change has a significant effect on the ferroelectricity of the smectic phases without causing any noticeable changes in the phase transition temperatures. We found that the polar 3F compound exhibits a previously not-reported phase sequence, where two polar smectic phases are formed from the paraelectric nematic phase. Moreover, the longer homolog exhibits paraelectric phases that uniquely stabilize polar phases in mixtures. The polarization current measurements and dielectric spectroscopy results reveal clear differences enabling the distinction of the $SmA_F$ and $SmC_F$ phases. By studying homologs exhibiting almost the same values and range of phase transitions, we also shed light on the coupling between the LC structure and their ferroelectric properties. The temperature and concentration range of polar phases occurrence in the mixture of the 3F and 4F compounds demonstrates the potential for creating multi-component LC materials with broad-temperature-range polar phases for the needs of modern photonics.

## Results and Discussion

The route of synthesis of compounds nF is presented in Scheme 1. Synthesis details are described in the ESI. The characteristic feature of the used method was protecting the carboxylic group in the 2-fluoro-4-hydroxybenzoic acid by the benzyl group from benzyl N,N'-dicyclohexylimidocarbamate (2), which is more selective and proceeds in mild conditions and higher yield forms the ester bond by directly treating dicyclohexylcarbodiimide. The needed benzyl N,N'-dicyclohexylimidocarbamate (2), was prepared from benzyl alcohol in a catalytic amount of cuprum chloride. The next step was an esterification reaction between benzyl 2-fluoro-4-hydroxybenzoate (3) and its proper 4-alkylbenzoyl chloride to obtain benzyl 2-fluoro-4-[(4-alkylbenzoyl)oxy]benzoate (4). Hydrogenolysis then released 2-fluoro-4-[(4-alkyl benzoyl)oxy]benzoic acid (5), which was then converted to the acid chloride using oxalyl chloride and then to the ester nF by reaction with 3,4,5-trifluorophenol.

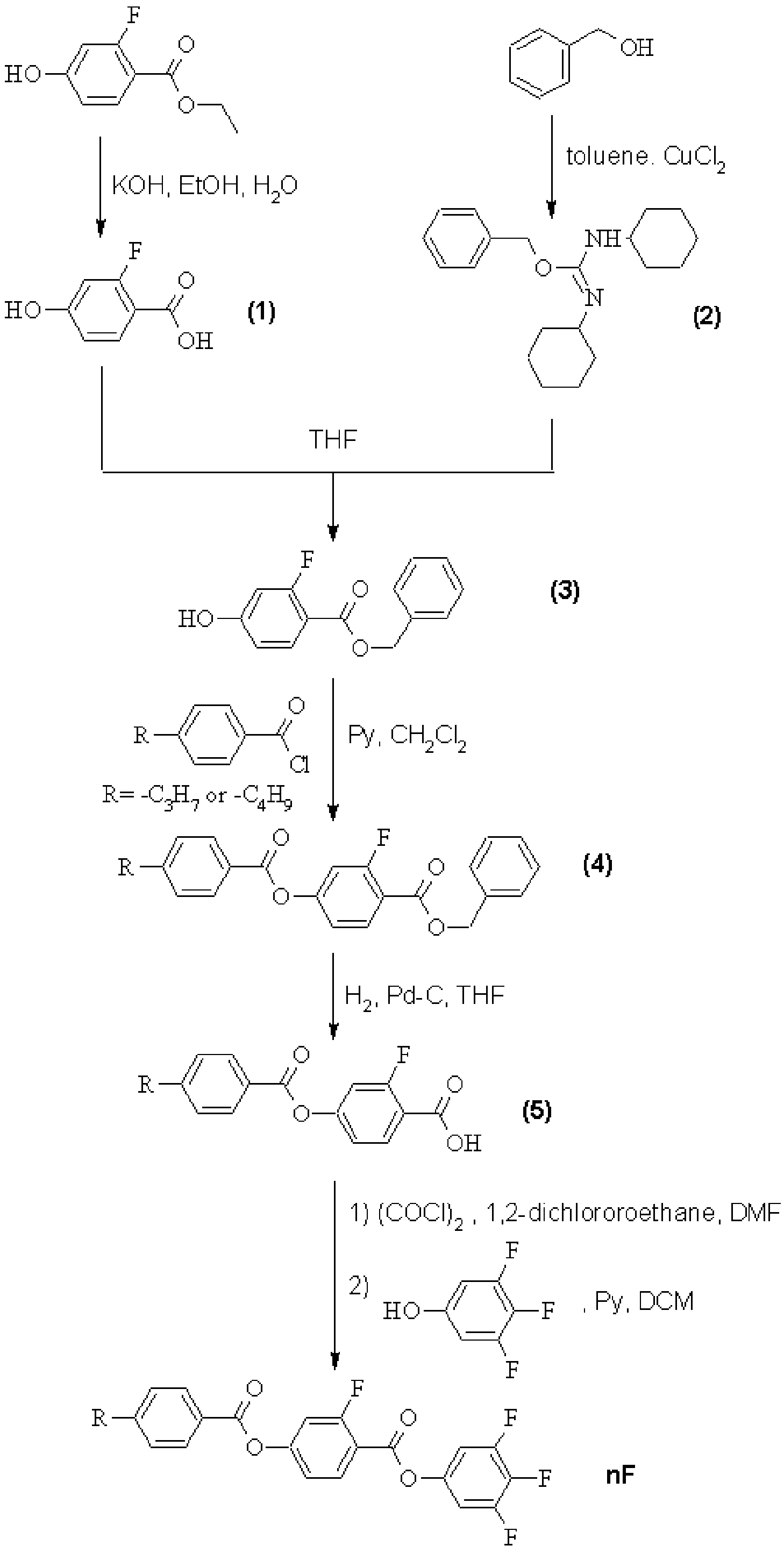

Scheme 1. The synthesis pathway of compounds nF, R is $C_3H_7$- for 3F or $C_4H_9$- for 4F.

After synthesizing and verifying the purity of the studied compounds, their mesomorphic properties were analyzed using polarized optical microscopy (POM) and differential scanning calorimetry (DSC). Table 1 presents the transition temperatures and corresponding changes in thermodynamic parameters for the 3F and 4F homologs. Detailed DSC traces and POM textures of studied compounds and mixtures with integrated peaks are given in Figures S1-2 and S4 (ESI), respectively.

Table 1. The phase transition temperatures [°C] (onset point), corresponding enthalpy changes [kJ mol$^{-1}$], in italic font, scaled entropy change, ΔS/R, for transition between the smectic and nematic phases [a.u.], as well as the temperature range of the nematic phase, N, and smectic phases, $SmA_F$ or SmA [°C] of the members of the homologous series nF, from DSC measurements determined during heating (upper rows) and cooling (down rows); values given in brackets were determined for monotropic phase.

| **Acronym** | **Cr** | **[°C]** ***[kJmol$^{-1}$]*** | **$SmC_F$** | **[°C]** ***[kJmol$^{-1}$]*** | **$SmA_F$** | **[°C]** ***[kJmol$^{-1}$]*** | **N** | **[°C]** ***[kJmol$^{-1}$]*** | **Iso** | **ΔS/R for N-$SmA_F$** | **N range [°C]** | **$SmA_F$ range [°C]** |
|---|---|---|---|---|---|---|---|---|---|---|---|---|
| | | 104.1 | | - | | 131.8 | | 137.1 | | 1.318 | 5,3 | 27,7 |
| | | *20.92* | | | | *4.44* | | *0.43* | | | | |
| **3F** | • | 89.4 | (•) | (99.1) | • | 130.5 | • | 136.2 | • | 1.311 | 5,7 | 31,4 |
| | | *-10.72* | | *(-0.03)* | | *-4.40* | | *-0.39* | | | | |
| **Acronym** | **Cr** | **[°C]** ***[kJmol$^{-1}$]*** | **$SmC_M$** | **[°C]** ***[kJmol$^{-1}$]*** | **SmA** | **[°C]** ***[kJmol$^{-1}$]*** | **N** | **[°C]** ***[kJmol$^{-1}$]*** | **Iso** | **ΔS/R for N-SmA** | **N range [°C]** | **SmA range [°C]** |
| | | 101.6 | | - | | 124.7 | | 130.5 | | 0,740 | 5,8 | 23,1 |
| **4F** | • | *26.88* | (•) | | • | *2.45* | • | *0.46* | • | | | |
| | | 84.2 | | (92.0) | | 123.2 | | 129.5 | | 0,686 | 6,3 | 31,2 |
| | | *-14.85* | | *(-0.12)* | | *-2.26* | | *-0.48* | | | | |

The liquid crystalline behavior of the homologs nF is characterized by notably similar phase transition temperatures and temperature ranges (Table 1). For compounds 3F and 4F, the melting and clearing temperatures differ by only 2.5 K and 6.6 K, respectively. These phase transitions also exhibit comparable enthalpy values. Furthermore, the temperature ranges of the nematic and orthogonal smectic phases are consistent, especially during the cooling process. However, the elongation of the alkyl terminal chain in nF series homologs by a single methylene group has a pronounced effect on the polar character of the smectic phases. The shorter homolog, 3F, exhibits the $SmA_F$ phase and the tilted $SmC_F$ phase with enantiotropic and monotropic character, respectively. In contrast, the 4F compound demonstrates an enantiotropic paraelectric SmA phase and the monotropic modulated smectic C phase ($SmC_M$). The entropy change scaled by the gas constant, ΔS/R, is a fundamental parameter to characterize changes in molecular ordering during phase transitions. For the 3F compound, the ΔS/R associated with the N- $SmA_F$ transition is roughly twice as large as the N-SmA transition for the 4F compound (Table 1). This difference likely reflects an additional entropic contribution due to dipole ordering in the $SmA_F$ phase compared to the conventional non-polar SmA phase. The phase sequence observed in compound 3F is particularly intriguing. Above the polar smectic phases, a well-defined, apolar nematic phase is present. A similar phase sequence has only recently been reported in the literature for a compound designated as EST-4.[25] However, in EST-4, the nematic phase exhibits cybotactic behavior with smectic short-range ordering. In contrast, the nematic phase of 3F shows no indications of local smectic ordering, as confirmed by X-ray diffraction studies (Figure S5, ESI). This finding underscores the uniqueness of the phase sequence in 3F, where a well-defined, apolar nematic phase transitions directly into a highly ordered polar smectic phase.

The POM textures for homologs in the nF series, shown in Figure 1A, further confirm the distinct nature of the smectic phases in the 3F and 4F compounds. During the transition from the nematic phase, which exhibits a typical schlieren texture, batonnet structures emerge. These batonnets show a single color during the transition to the SmA phase and display a rainbow effect when transitioning to the $SmA_F$ phase (Figure 1Ah and 1Ad, respectively). The textures of orthogonal smectic phases also differ between the homologs studied. The SmA phase of 4F displays the commonly observed focal-conic fan texture, whereas the $SmA_F$ phase of 3F exhibits a distinctive mosaic-like texture characterized by sharp boundaries and rainbow-like elements along these boundaries (Figure 1Ac). A similar POM texture of the $SmA_F$ phase, observed between untreated glass plates, was presented in other works.[19,20] Upon further cooling to the monotropic tilted phases, the focal-conic fan and mosaic-like textures become broken and banded in both homologs of the nF series. This effect is particularly pronounced during the transition to the $SmC_F$ phase of the 3F compound. Textures observed in SmA and SmC phases for 4F homolog are similar to typical textures observed for nonchiral smectics.[26]

To elucidate the loss of smectic phase polarity following chain elongation in the studied homologous series, quantum chemistry calculations were performed using the B3LYP/6-311G+(d,p) level of density functional theory (DFT). The optimized geometry, along with Cartesian axes, dipole moment vector, and electronic potential surface (ESP) for the 3F and 4F homologs are shown in Figure 1B. Table 2 collects the values of the dipole moments and selected geometry parameters for these compounds. Higher length-to-width ratio for the 4F molecule compared to 3F would be expected to lead to notable changes in phase transition temperatures and an increase in smectogenity, as shown for other known homologs with polar phases. [15,16,24,27] Surprisingly, these effects were not observed in the homologs studied in this work. Both 3F and 4F compounds exhibit similar dipole moment values (Table 2) and almost the same ordering of their vectors relative to the long molecular axis (Figure S3, ESI). Compared to previously presented analogs (the nCN series)[15] which differ only by having a terminal CN polar group instead of an F atom, the dipole moment values of the nF homologs are nearly 20% lower. Based on the commonly observed trend of increased smectic tendencies with decreasing dipole moment,[19,23,24] this is likely the reason for the emergence of smectic phases in nF compounds rather than the nematic phases seen in nCN analogs. Furthermore, compared to 3CN,[15] compound 3F exhibits a similar longitudinal wave of surface charge density along the long molecular axis, which promotes polarity in LCs,[28] owing to the identical rigid core structure. This characteristic enables the preservation of phase polarity below the high-temperature apolar nematic phase. On the other hand, Madhusudana suggests that elongation of the alkyl terminal chain increases charge density at one end of the molecule, contributing to the formation of the anti-parallel ordering of molecules to minimize dipolar energy.[28] This, in turn, contributes to the formation of non-polar phases, which explains the observed lack of ferroelectricity in the 4F compound.

To further investigate the distinct nature of the smectic and nematic phases in the studied compounds, and to deeply evaluate the balance between paraelectric and ferroelectric characteristics, a binary mixture system comprising 3F and 4F was prepared and studied (Figure 1C, D). Ferroelectric orthogonal and tilted smectic phases persist in mixtures containing up to a 0.75-mole fraction of compound 4F. The direct N–$SmA_F$ phase transition is absent only at or above the equimolar composition of the 3F and 4F mixtures. All phase transitions in the pure compounds (Figure S1, ESI) and their mixtures (Figure 1D and S2) are of the first order, as evidenced by peaks in the DSC tracers for each transition. The characteristic textures, as described above for the pure components, observed for individual mixtures at various temperatures, further confirm the types and sequences of liquid crystalline phases in the 4F-3F system (Figure S4, ESI). Such a strong tendency to sustain ferroelectric phases in a binary mixture of a polar and an apolar compound closely resembles the scenario described in [29]. In that study, the authors demonstrated an unexpected enhancement in the stability of the polar nematic phase within a mixture of two diastereomers, cis- and trans-DIO. The cis-DIO compound, a polar crystalline material, maintained the NF phase in a mixture with trans-DIO even beyond the equimolar composition. This phenomenon was attributed to the formation of cybotactic nano-clusters in the ferroelectric phase, driven by molecules with similar dipole moments and chemical structures. A similar mechanism likely governs the stabilization of polar smectic phases in the studied homologous mixture 4F-3F. Importantly, a key advantage of the observed balance between ferroelectricity and paraelectricity in the 4F–3F system is the remarkable thermal stability of such mixtures. As recently demonstrated, the 1,3-dioxane ring of DIO exhibits thermal instability, which may indicate potential instability in mixtures based on different diastereomers of DIO.[23] In contrast, the 4F–3F system appears to offer enhanced thermal robustness, as it consists of two stable homologs.

Table 2. Components and the total dipole moment values as well as length (L), width (D), and length-to-width ratio (L/D) calculated for the optimized geometry of 3F and 4F compounds.

| **Acronym** | **Dipole moment [Debye]** | | | | **Geometrical parameters** | | |
|---|---|---|---|---|---|---|---|
| | $\mu_x$ | $\mu_y$ | $\mu_z$ | $\mu_{total}$ | **L [Å]** | **D [Å]** | **L/D** |
| **3F** | 5.52 | -0.29 | 9.91 | 11.35 | 19.69 | 5.95 | 3.31 |
| **4F** | 5.53 | -0.29 | 9.99 | 11.43 | 20.95 | 5.97 | 3.51 |

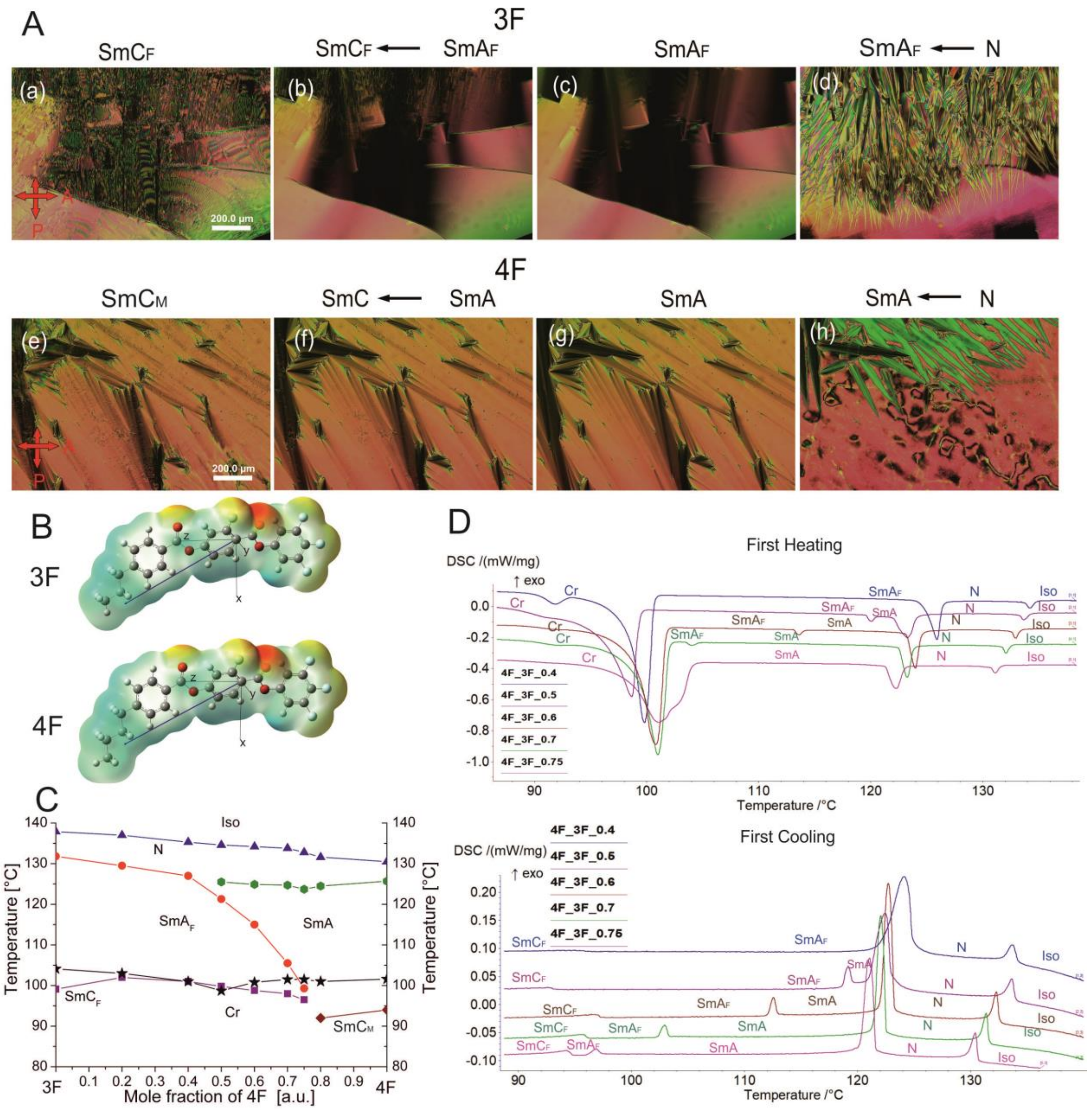

Figure 1. Mespomorphic and electrostatic properties of the 3F and 4F compounds, and their mixtures. (A) POM images, obtained for material between untreated glass plates on cooling from the isotropic phase, of 3F (a-d), and 4F (e-h), under crossed polarizes. (B) Optimized geometries of 3F and 4F at B3LYP/6-311G+(d,p) level of DFT, with the electrostatic potential (ESP) on the 0.0004 au electron density isosurface. (C) Phase diagram obtained on heating for the bicomponent mixture systems 4F-3F. (D) Stacked DSC heating (top) and cooling (bottom) traces for 4F-3F binary mixtures, labeled as the mole fraction of the 4F component.

To confirm the phase identification and reveal their structure, X-ray diffraction (XRD) experiments have been performed. In the nematic phase, only broad diffuse scattering signals were registered at both: small and high-angle ranges (Figure S5, ESI), pointing to a lack of the long-range positional order characteristic of this phase. In smectic phases of the compound 3F, the width of the low angle diffraction signal narrows, to become limited by instrumental broadening, evidencing the appearance of long-range positional order of molecules. The smectic layer spacing, *d*, determined from the position of the signal, corresponds to the molecular length and it was found to be almost temperature independent within the temperature range of the $SmA_F$ phase (Figure 2A, B). In the $SmC_F$ (Figure 2C) phase the tilting of the molecules with respect to the layer normal leads to the gradual decrease of the layer thickness, by ~0.05 Å $K^{-1}$. The tilt angle estimated from the changes in *d*-spacing reaches ~15 deg. 10 K below the $SmA_F$-$SmC_F$ phase transition. In both smectic phases, the high-angle scattering signal is diffused, consistent with liquid-like correlations of molecular positions within the layers. Interestingly, the signal is clearly split into two maxima, corresponding to periodicities of 4.6 and 3.5 Å (Figure S6, ESI); such a splitting might be ascribed to local biaxiality resulting from partial inhibition of molecular rotation about the long axis.

XRD experiments performed for compound 4F (Figure 2D) revealed similar phase behavior as for 3F, however with two noticeable differences: (*i*) layer spacing in orthogonal, SmA phase gradually expands on heating, with thermal expansion coefficient 0.004 Å $K^{-1}$ and (*ii*) the structure of tilted smectic phase is not simple lamellar. The lower temperature smectic phase shows addition density modulation along the layers, as evidenced by weak low-angle diffraction signals at the equatorial direction and satellite signals accompanying the main diffraction signal due to layer periodicity (Figure 2E, F), thus the phase is referred to as modulated SmC ($SmC_M$) phase. Positions of additional diffraction signal correspond to a periodicity of about 115 Å however, it was not possible to unambiguously determine the symmetry of the 2D crystallographic lattice. Both SmA and $SmC_M$ of the compound 4F exhibit a liquid-like order of neighboring molecules within the layers.

The ferroelectric (non-centrosymmetric) character of both smectic phases of the 3F compound has confirmed their second harmonic generation (SHG) activity. In the experiment, the cell was illuminated using the IR laser ($\lambda$ = 1064 nm) and the SHG signal with doubled frequency (green color) started to be observed at the N-$SmA_F$ (Figure 2G) phase transition and it was visible throughout the entire temperature range of $SmA_F$ and $SmC_F$ (Figure 2H and 2I).

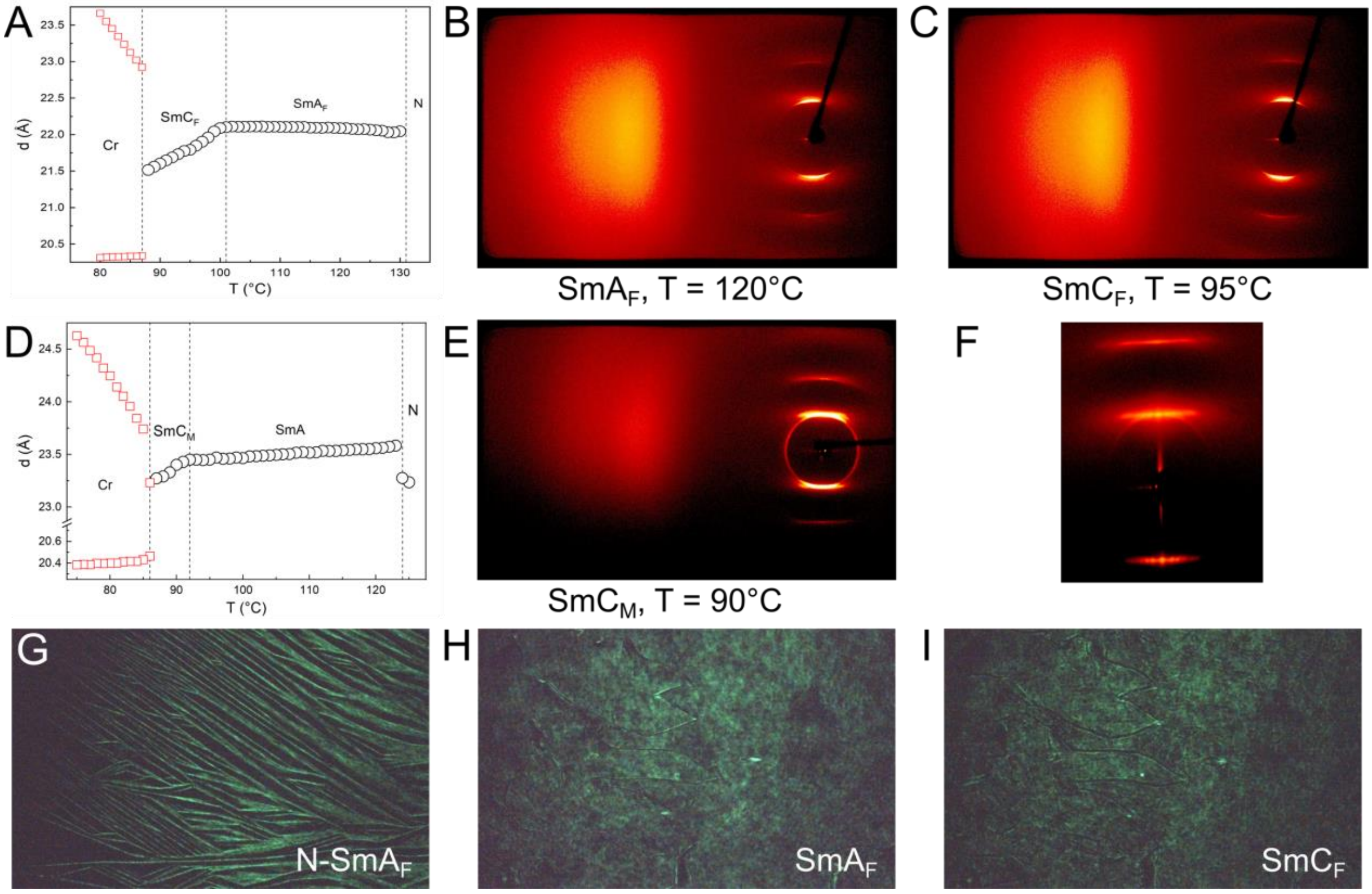

Figure 2. (A) Layer spacing (*d*) of the 3F compound, as a function of temperature. The data were obtained from X-ray diffraction measurements on cooling from the nematic (N) phase through the smectic phases (black circles) to the crystal (Cr) phase (red squares). (B, C) 2D XRD patterns in the SmA$_F$ (120°C) and SmC$_F$ (95°C) phases of 3F. (D) Layer spacing determined in LC phases of the 4F compound (black circles), periodicities corresponding to the signals recorded in the Cr phase are shown as red squares. (E, F) 2D XRD patterns of the modulated smectic (SmC$_M$) phase. (G, H, I) The SHG microscopy images taken at the N-SmA$_F$ phase transition and in the SmA$_F$ and SmC$_F$ phases.

Figure 3A demonstrates the current response of the 3F compound under the applied AC electric field in two different polar phases. The ferroelectric switching at 120°C, SmA$_F$ phase, followed the triangular-shaped electric field with two characteristic peaks. These peaks correspond to two different times of molecular switch. Usually, the twin polarization reversal peaks in a short time domain are evidence of the antiferroelectricity of the polar liquid crystalline material. In the well-known antiferroelectric smectic SmC$_A$*phase, the first peak in the time scale is related to the molecular switching within one layer. While the switching mechanism in the adjacent layer is responsible for the appearance of the second peak. The polarization charge in such material is evenly distributed between the two peaks. Both peaks are the same and do not evolve with temperature, one peak does not increase its area at the expense of the other. In our case, the SHG experiment confirms that both polar smectic phases exhibit ferroelectric behavior. Moreover, the spontaneous polarization peaks change their shape independently of each other with temperature (Figure 3B). Therefore, the presented ferroelectric polarization switching in $SmA_F$ must involve overcoming two potential barriers with lower and higher energy. The two-step polarization switching process in $SmA_F$ is related to the nematic-smectic order transition. The low electric field mechanism is connected to the nematic reorientation of dipoles in the direction of the electric field vector. This is evidenced by the occurrence of a molecular mode typical for the N phase observed by dielectric spectroscopy in the ferroelectric smectic A phase. This issue will be discussed later. The peak decreases during cooling and disappears upon entering the $SmC_F$ phase. The first process is immediately followed by the second one, in which the polarization aligns in smectic layers when the high electric field is applied.

In the $SmC_F$ phase, the dipole reversal process is also visible by two peaks. However, in this case, the smectic switching mechanism (the polarization switching within the layers) is detected as a main, high current peak, observed at high electric fields. The additional much smaller peak is triggered at the lower electric fields. A similar low energy peak has been already observed in newly synthesized polar SmC compounds[17,24] and was linked to the inducing of the molecular tilt within the $SmC_F$ phase.[22]

The 3F compound has a maximum $P_S$ near the N-$SmA_F$ phase transition (Figure 3C). The spontaneous polarization in the investigated ferroelectric smectic compound is in the same range (~ 6 $\mu C/cm^2$) as reported for the compounds with the ferroelectric nematic phase.[12,13] As in the chiral ferroelectric smectics, the spontaneous polarization depends on the tilt angle.[3] In comparison to the $SmA_F$ phase, in $SmC_F$ the value of $P_S$ decreases around 1 $\mu C/cm^2$, which is still tens of times larger than for typical SmC*.[3]

Due to the extensive discussion on the high capacity contribution of the polyimide aligning layers and cell thickness to the dielectric response results, the measurements were performed in cells of constant thickness with bare electrodes.[30–34] Nevertheless, the obtained results describe the electrical properties of ferroelectric smectics in a qualitative rather than quantitative way. The dielectric spectra of the 3F compound give clear evidence of two polar liquid crystalline phases, where we observe a significant increase in the real part of the electric permittivity $\varepsilon'$ (Figure 3D). In 4F, we observe a classical paraelectric Iso-N-SmA phase transition (Figure 3E). Only in the modulated smectic phase, there is a slight $\varepsilon'$ increase, most likely related to the creation of a higher-ordered smectic.

The dielectric response in the paraelectric nematic phase of 3F is very weak. We observe only at the isotropic-nematic phase transition a slight increase of electric permittivity due to the induced homeotropic alignment of highly polar molecules on the glass covered by ITO. Next, at the N-$SmA_F$ transition the rapid increase of $\varepsilon'$ is noticed. For this phenomenon is responsible for the strong and low-frequency dielectric mode visible in the spectrum of imaginary part of electric permittivity $\varepsilon''$ (Figure 3F). This mode is detectable only at the narrow temperature range of $SmA_F$. The disappearance of this mode causes a visible decrease in electric permittivity between two polar phases. Despite the clear decrease in $\varepsilon'$ (temperature range: 122°C – 100°C), the ferroelectric behavior of the smectic A phase remains preserved. The dielectric loss spectrum exhibits, besides the low-frequency relaxation, an additional high-frequency absorption peak. The second dielectric mode softens with temperature decreasing towards $SmC_F$. In the polar $SmC_F$ phase, only a strong single relaxation is visible, which gives a greater contribution to the real part of electric permittivity than the two modes observed in $SmA_F$.

The temperature dependence of the relaxation frequency $f_R$ and the dielectric strength $\delta\varepsilon$ in the different phases are depicted in Figure 3G and 3H, respectively. In the N phase, $f_R$ exhibits the expected slowing down behavior with the temperature decreasing, while $\delta\varepsilon$ increases. This temperature dependence is typical for a non-collective molecular relaxation around a short molecular axis (S-mode). This mode is continued in the $SmA_F$ but only at the beginning of the phase, for the first 11 degrees, and it is responsible for the noticeable peak of electric permittivity in 3D plots (from 134°C to 123°C). The dielectric strength of the second dielectric mode in $SmA_F$ increases and the relaxation frequency decreases with temperature decreasing. This behavior is characteristic of the soft mode.[35] A similar relationship was observed at the transition between paraelectric SmA* - ferroelectric SmC* and SmA* - antiferroelectric $SmC_A$* in chiral smectics.[36,37] Therefore, this mode can be described as a collective amplitude mode. At the onset of the ferroelectric smectic C phase, $f_R$ and $\delta\varepsilon$ become almost constant with temperature. This relationship indicates the collective nature of the relaxation. Considering the symmetry of the $SmC_F$ the observed dielectric mode can be an analogy to the Goldstone mode (GM) in SmC*.[35] Tilting of molecules by an $\theta$ angle in $SmC_F$ causes the disappearance of amplitude mode (soft mode) and the appearance of the phason mode (GM) (Figure 3I). In the weak AC electric field, we detect the fluctuation of the azimuthal $\phi$ angle. Of course, in a perfectly aligned homeotropic $SmC_F$, the GM mode would not be electrically active. However, the XRD studies show that 3F compounds exhibit imperfect orientational and transitional order. Therefore, GM is detectable in in tens of kilohertz frequency regime. The change in polarization associated with a change in the angle $\phi$ is much larger than the change in polarization associated with a change in the $\theta$ tilt angle. For this reason, the calculated dielectric strength of the Goldstone mode is much larger than the soft mode. It is worth mentioning that in both compounds a parasitic ionic relaxation was noticed. Its contribution to the dielectric response was taken into consideration in molecular calculations.

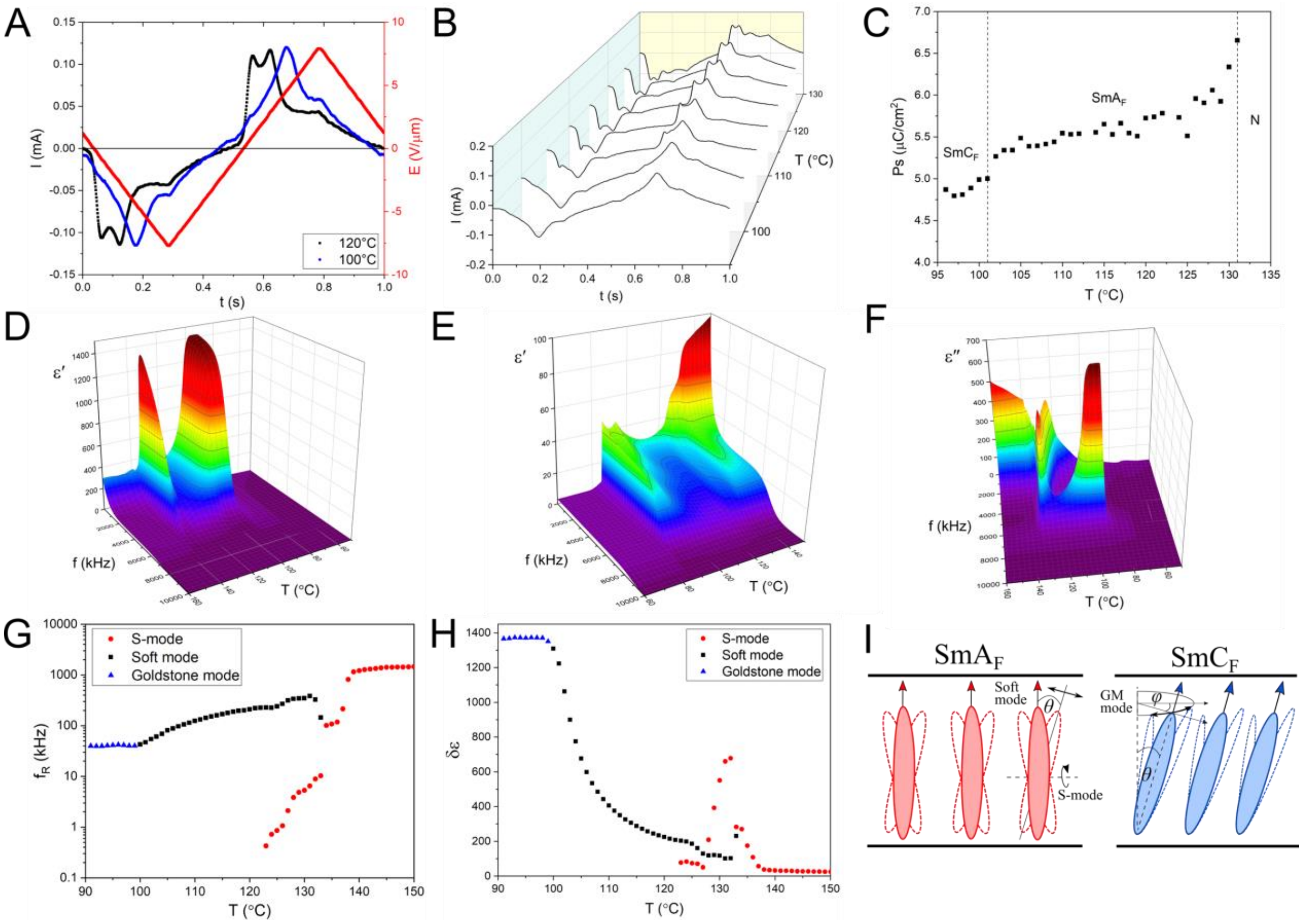

Figure 3. Electrical properties. (A) Current response ($I$) to a triangular electric field (red line) obtained in the $SmA_F$ phase (120°C) and $SmC_F$ phase (100°C) versus time ($t$). (B) Three-dimensional (3D) dependence of current peaks in polar smectic phases. (C) Spontaneous polarization $P_S$ values obtained on cooling as a function of temperature. (D, E) 3D dielectric spectra of the real part $\varepsilon'$ of electric permittivity measured in 3F and 4F compounds. (F) 3D dielectric loss spectra in 3F. (G) Temperature dependence of relaxation frequency $f_R$ and (H) dielectric strength $\delta\varepsilon$ of detected dielectric modes in 3F. (I) Different relaxation mechanisms are proposed for a non-chiral ferroelectric smectic A and C phase.

## Conclusions

To conclude, we reported on the first experimental observation of the formation of the two ferroelectric layer fluid, with orthogonal and tilted molecules, from the paraelectric, well-defined nematic phase. The shorter homolog (3F) of the investigated materials is characterized by a direct $SmA_F$ – $SmC_F$ phase transition, while the compound with four carbons in the methylene unit (4F) is characterized by a strong affinity for stabilizing the polar structure. This work presents noticeable differences, in optical and electrical properties resulting from the layered ferroelectric ordering and mechanism of molecular tilt in the non-chiral liquid crystalline systems. The emergence of the $SmA_F$ phase is associated with the observation of two neighboring peaks of the polarization current and the soft-type dielectric mode. The tilted $SmC_F$ phase is characterized by a main high current peak, observed at high electric fields, and a smaller peak at lower electric fields. Only one strong collective mode defined as the Goldstone-type mode is detected in $SmC_F$.

In the studied homologs, the primary determinant for the occurrence of polar phases in liquid crystal compounds is not the change in the molecule's length-to-width ratio. Instead, it is the alteration in charge density difference at the molecular ends, resulting from chain elongation, that plays this key role. The 3F compound has too small a dipole moment and charge density oscillation across a long molecular axis to create the ferroelectric nematic phase but is large enough for ferroelectricity to appear in smectic A ($SmA_F$) and smectic C ($SmC_F$). Furthermore, it was observed that polar smectic phases appear in mixtures even with a threefold excess of the non-polar 4F relative to the polar 3F. This indicates a strong tendency of compounds in the nF homologous series to maintain ferroelectric properties in smectic phases. This is likely due to the same modulation of the electrostatic potential along the rigid core of the molecules in both homologs. The above findings and conclusions provide a clear pathway on how to carefully design new non-chiral LCs for formulating multi-component mixtures with ferroelectric properties, high thermal stability, and significant application potential.

## Author Contributions

DW synthesized the compounds and wrote the corresponding parts of the draft, MC prepared bicomponent mixtures, performed optical microscopy and DSC measurements, molecular modeling and wrote the original draft, RD synthesized the compounds, PP analyzed the data, JS measured the SHG signal, DP performed XRD measurements analyzed the data, and wrote the corresponding parts of the draft. MM performed spontaneous polarization and dielectric spectroscopy measurements, analyzed the data, directed the project, and wrote the original draft. All authors contributed to the scientific discussions.

## Conflicts of interest

There are no conflicts to declare.

## Acknowledgments

The authors acknowledge the financial support from the National Science Centre in Poland project no. 2022/06/X/ST5/01316 and 2024/53/B/ST5/03275, the MUT University grant UGB 531-000030-W900-22. The authors thank Monika Zając for performing GC/HPLC–MS analysis and Jakub Herman for performing $^{1}$HNMR and $^{13}$CNMR spectra measurements.

*Supplemental Information*

# The balance between paraelectricity and ferroelectricity in non-chiral smectic homologs

Dorota Węgłowska, Michał Czerwiński, Robert Dzienisiewicz, Paweł Perkowski, Jadwiga Szydłowska, Damian Pociecha, Mateusz Mrukiewicz*

Dorota Węgłowska, Michał Czerwiński, Robert Dzienisiewicz

*Institute of Chemistry, Military University of Technology, Kaliskiego 2, 00-908 Warsaw, Poland*

Paweł Perkowski, Mateusz Mrukiewicz

*Institute of Applied Physics, Military University of Technology, Kaliskiego 2, 00-908 Warsaw, Poland*

Damian Pociecha, Jadwiga Szydłowska

*Faculty of Chemistry, University of Warsaw, Żwirki i Wigury 101, 02-089 Warsaw, Poland*

Corresponding author: mateusz.mrukiewicz@wat.edu.pl

Contents:

1. Experimental methods

*Differential scanning calorimetry*

Differential scanning calorimetry (DSC) measurements were carried out using a Netzsch DSC 204 F1 Phoenix calorimeter, which was calibrated with indium, zinc, and water standards. The heating and cooling processes were performed at a rate of 2.0 K/min, and the samples were maintained in the aluminum crucibles under a nitrogen atmosphere with a gas flow rate of 20.0 ml/min. The transition temperatures and corresponding enthalpy change values were determined from the heating and cooling curves.

*Polarizing optical microscopy*

Optical investigations were performed with an OLYMPUS BX51 polarized optical microscope equipped with a Linkam TMS93 temperature controller and a THMSE 600 heating stage. The samples were placed between untreated glass plates to observe their "natural" textures.

*Molecular modeling*

The dipole moment values for the optimized molecular geometry were computed using the GAUSSIAN 09 molecular simulation software.[1] Structural optimization and other related calculations were performed using the B3LYP hybrid functional in conjunction with the 6-311G+(d,p) basis set, as previously described in our earlier work.[2] A frequency analysis was conducted to ensure that the obtained conformation represented a true energy minimum.

*X-ray diffraction*

X-ray diffraction (XRD) studies in broad diffraction angle range were performed with a Bruker GADDS system. The system is equipped with a microfocus type X-ray tube with Cu anode and Vantec 2000 area detector. For small angle XRD experiments a Bruker Nanostar system was used with a microfocus X-ray tube with Cu anode, MRI heating stage, and Vantec 2000 area detector. In both cases samples were prepared as a droplets/thin films on a heated surface.

*Second harmonic generation*

In the Second Harmonic Generation (SHG) experiment, a solid-state infrared laser EKSPLA NL202 was used. Laser radiation with a wavelength of $\lambda$ = 1064 nm incident perpendicularly onto a planarly aligned cell of thickness 5 μm. To avoid material decomposition, the pulse energy was adjusted. The laser generated the 9 ns pulses at a 10 Hz repetition rate with a maximum energy of 2 mJ. At the entrance to the cell, an IR filter was installed on the setup, whereas a green filter was placed at the output of the SHG signal from the sample.

*Spontaneous electric polarization*

The triangle-wave technique was used to induce the repolarization current. The electric signal of frequency 1 Hz was applied by a Hewlett Packard 33120A waveform generator and gained using an FLC Electronics F20ADI voltage amplifier. The voltage drop across a 10 kΩ resistor was registered on a Hewlett Packard 54601B oscilloscope. To calculate the spontaneous polarization $P_S$ value, the peak current was integrated in the time domain using the following formula:

$$P_S = \int \frac{I_P}{2S} dt \quad (1)$$

where $S$ is the active electrode area, $I_P$ is the polarization current flowing through the cell after the separation of the capacitor charging current and the ionic current. The temperature was stabilized using a Linkam TMS 93 temperature controller with a THMSE 600 heating stage. The measurements were done in 5 μm thick cell without any aligning layers.

*Dielectric Spectroscopy*

The complex electric permittivity $\varepsilon^*$ was measured at low (0.1 V) signal voltage over the frequency $f$ range from 100 Hz to 10 MHz by using a Hewlett Packard 4294A impedance analyzer. The dielectric studies were performed in 5 μm thick cells with low resistivity (10 Ω/sq) ITO electrodes. Cells without alignment polyimide layers were used. The temperature was stabilized with 0.1°C accuracy using a Linkam TMS 92 temperature controller with a THMSE 600 heating stage. The calculated real $\varepsilon'$ and imaginary $\varepsilon''$ parts of electric permittivity were fitted to the Cole-Cole model[3] with the ionic contribution to determine the dielectric strength $\delta\varepsilon_i$ and characteristic relaxation frequency $f_{R_i}$ of the $i^{\mathrm{th}}$ relaxation mode:

$$\varepsilon^* = \varepsilon_\infty + \sum_i \frac{\delta\varepsilon_i}{1+\left(j\frac{f}{f_{R_i}}\right)^{1-\alpha_i}} - j\frac{\sigma}{2\pi f \varepsilon_0}, \quad (2)$$

where $\varepsilon_0$ is the electric permittivity of free space, $\varepsilon_\infty$ is the high-frequency limit of permittivity, $\alpha$ is the distribution parameter, and $\sigma$ represents conductivity due to the electrode polarization process visible in dielectric spectra at low frequencies (below 1 kHz).

2. Supplementary results

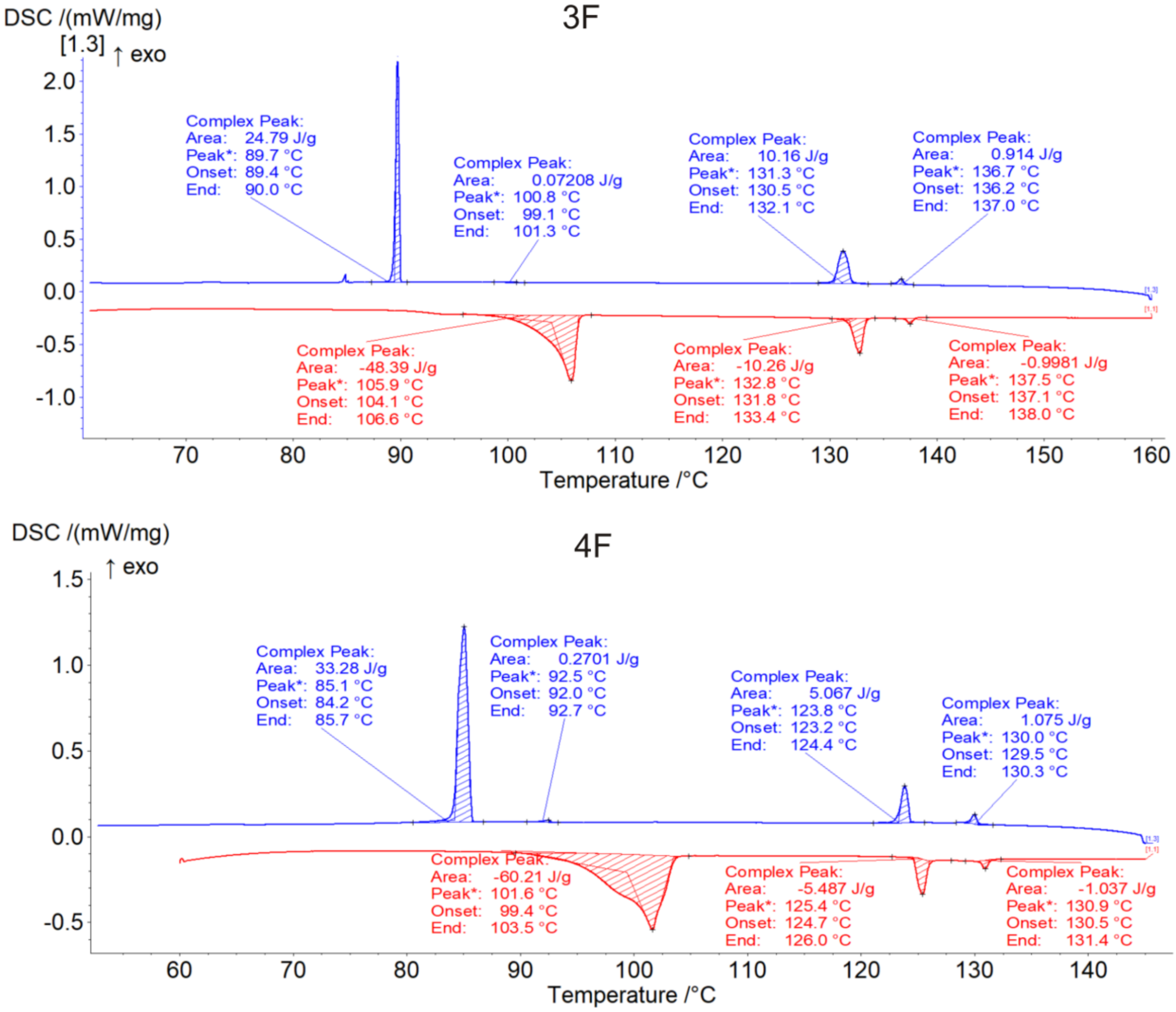

Figure S1. The DSC tracers of compounds 3F and 4F in the heating cycles (down curves) and cooling cycles (upper curves).

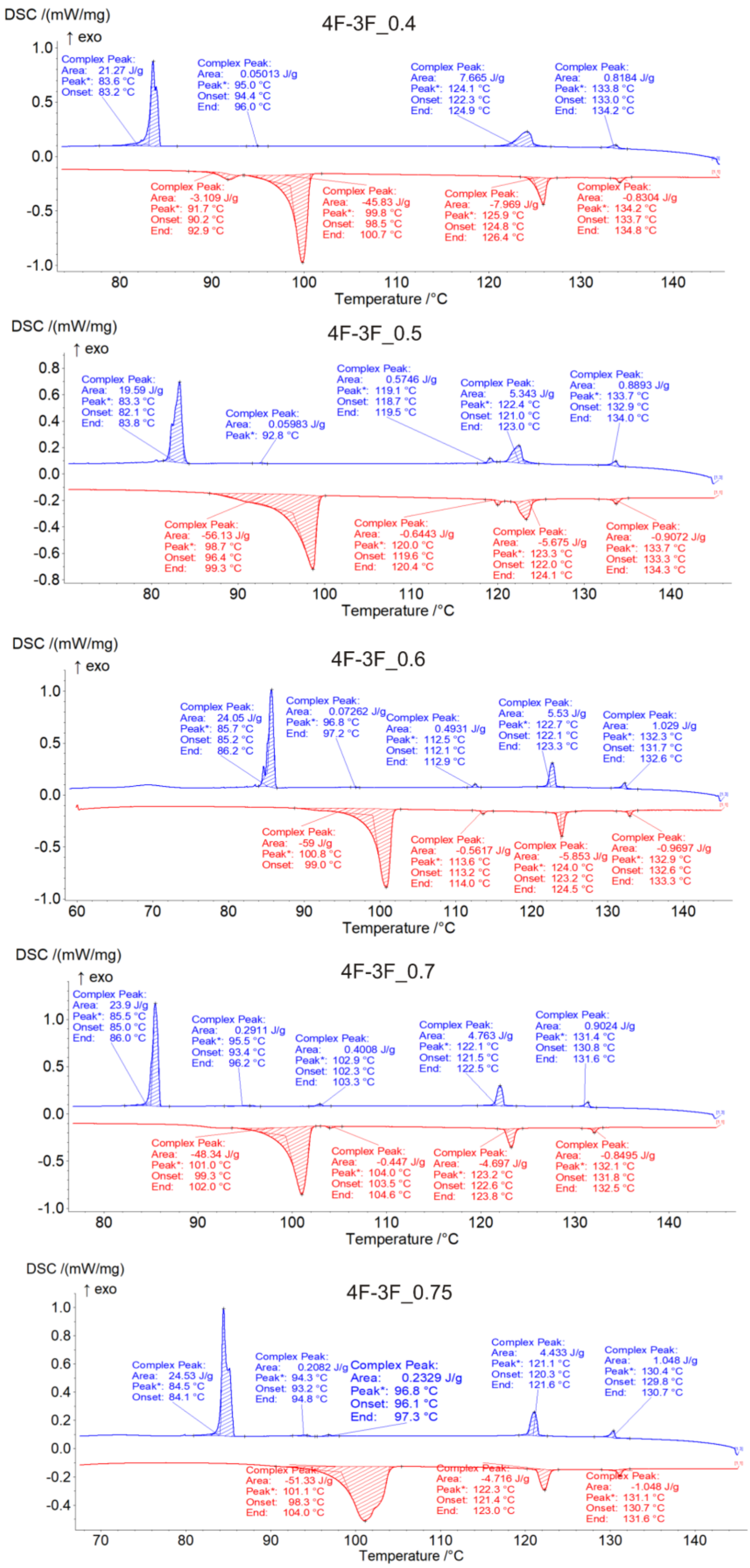

Figure S2. The DSC tracers of mixtures 4F-3F in the heating cycles (down curves) and cooling cycles (upper curves).

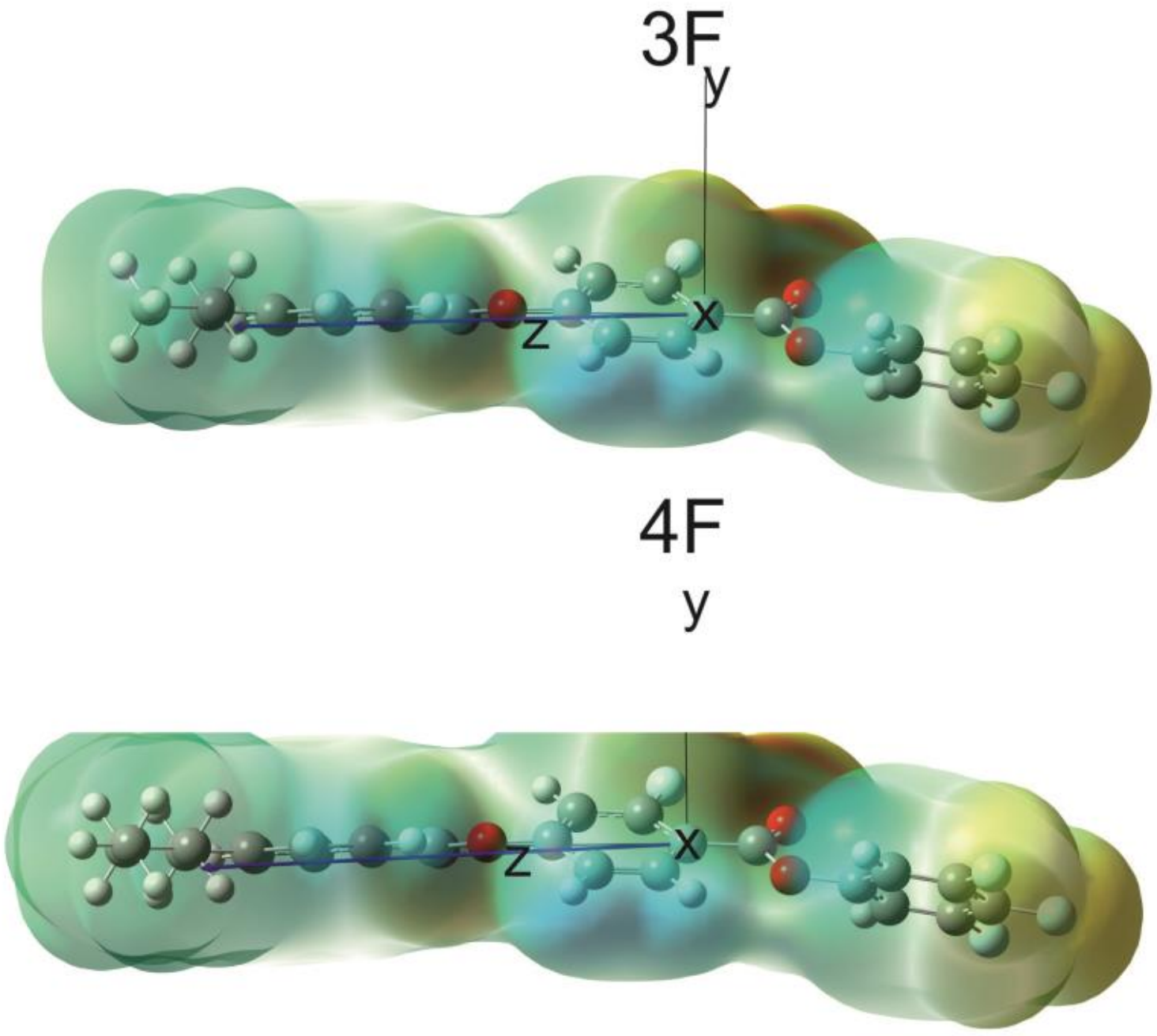

Figure S3. The optimized geometric general structure of compounds 3F and 4F aligned along the principal molecular axis with a blue arrow showing the direction of the molecular dipole moment; whereas *x* is the axis in the plane of the phenyl rings, *y* - is the axis perpendicular to the plane of the phenyl rings and *z*-the axis along to the principal molecular axis.

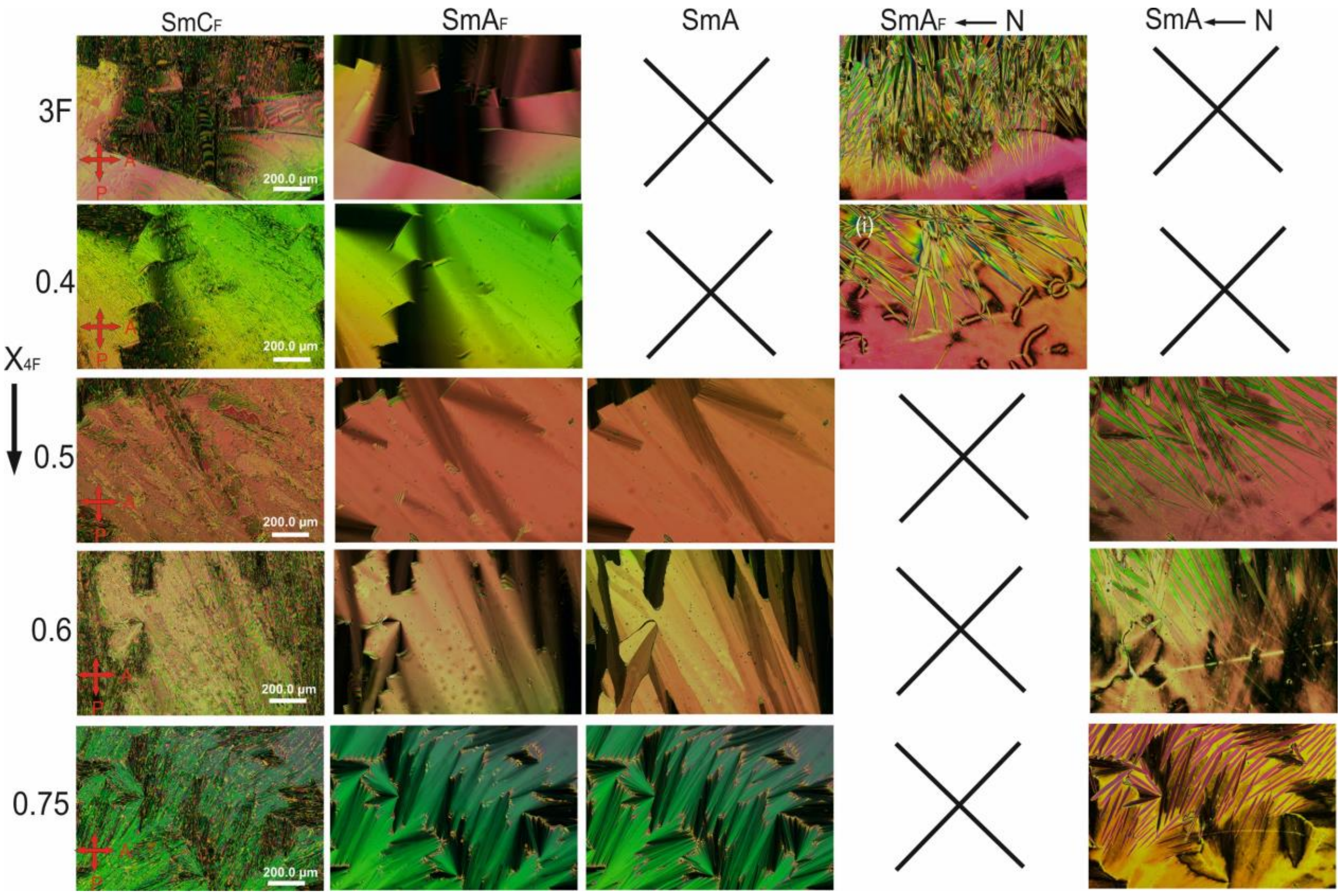

Figure S4. The POM texture evolution of 3F compound and mixtures 4F-3F system during the cooling process under crossed polarizers for LC filled between untreated glass plates.

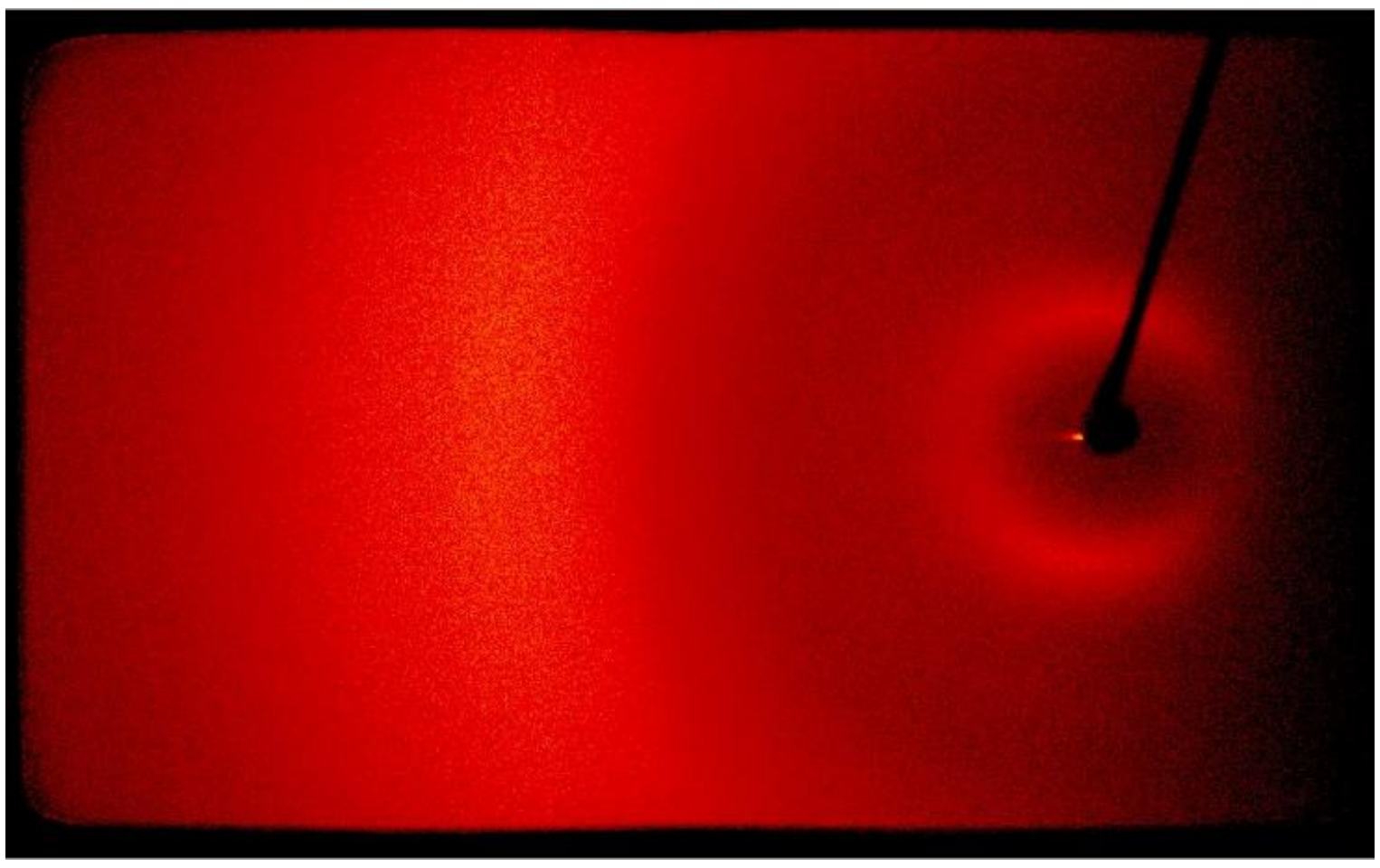

Figure S5. XRD diffraction pattern in the nematic phase (135°C) of 3F.

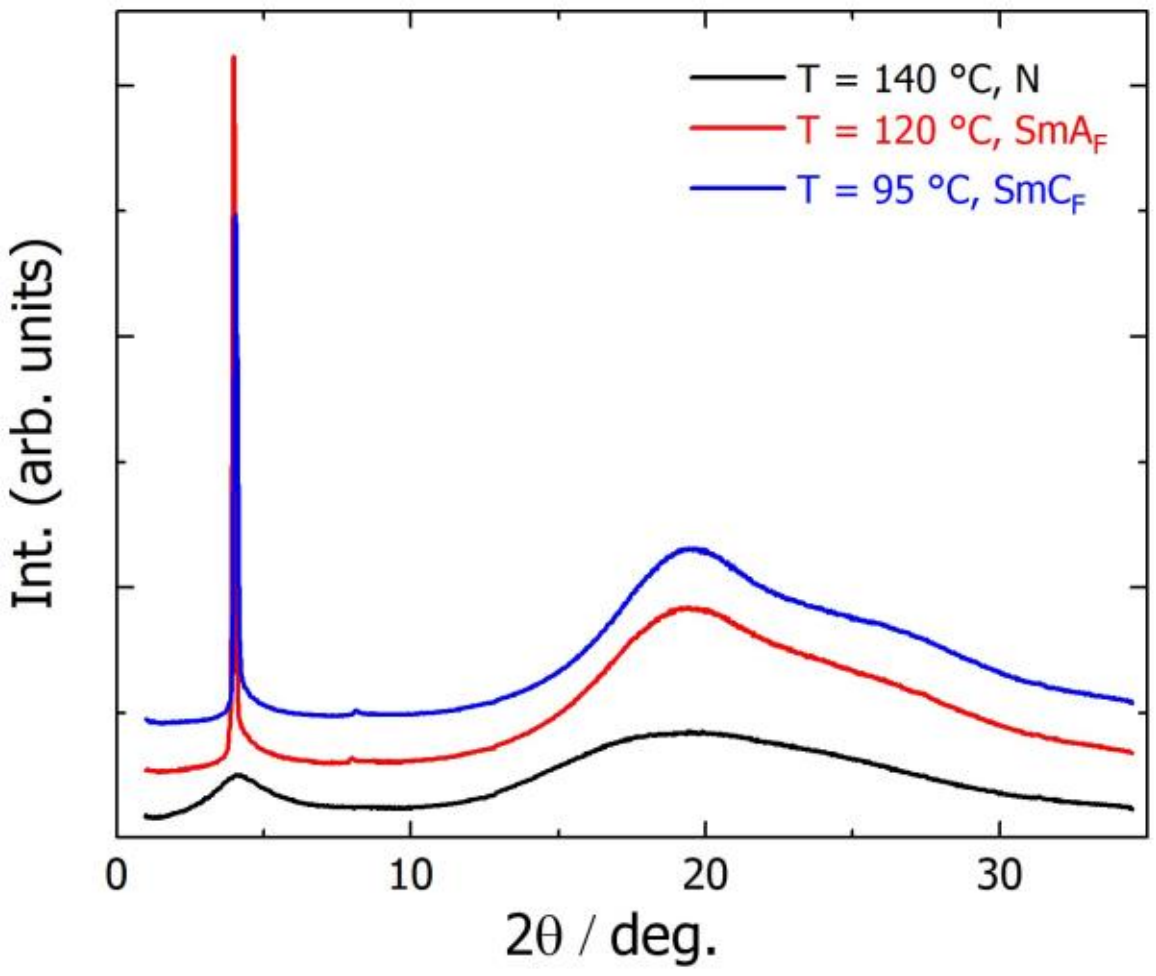

Figure S6. Broad-angle XRD diffractograms recorded in LC phases of compound 3F.

3. Chemical synthesis and characterization

**Preparative procedures**

The purity of intermediates and the main compounds were determined by thin layer chromatography (TLC), GC-MS(EI) (Agilent 6890N, Santa Clara, CA, USA), and HPLCPDAMS (API-ESI) (Shimadzu Prominence LC20) chromatography systems. The structures of the final compounds were confirmed by mass spectra registered on the HPLC chromatography system with a diode array detector (SPD-M20A) and mass detector (LCMS-85 2010EV) and by $^{1}$H and $^{1}$C NMR spectroscopy (Bruker, Avance III HD, 500 Hz; $CDCl_3$, Billerica, MA, USA).

**Synthetic procedures** are described for an exemplary compound **3F**.

*2-fluoro-4-hydroxybenzoic acid* **(1)**

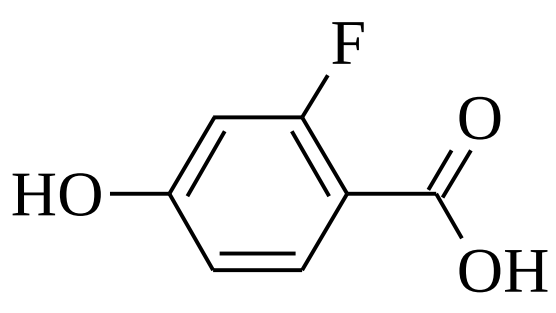

A mixture of ethyl 2-fluoro-4-hydroxy benzoate (15.0 g, 81 mmol), potassium hydroxide (18.8 ml, 334 mol) in water (25 ml) and ethanol (100 ml) was refluxed for 6 hours. When the reaction was complete (TLC), the reaction mixture was poured into 2M HCl (100 ml), stirred, and concentrated at low pressure. The crude product crystallized from water (100 ml). Yield: 10.6 g (85%); $M_p$=195–197°C. GC: 99.8 %; MS(EI) m/z: 156 $[M+H]^+$

*Benzyl N, N'-dicyclohexylimidocarbamate* **(2)**

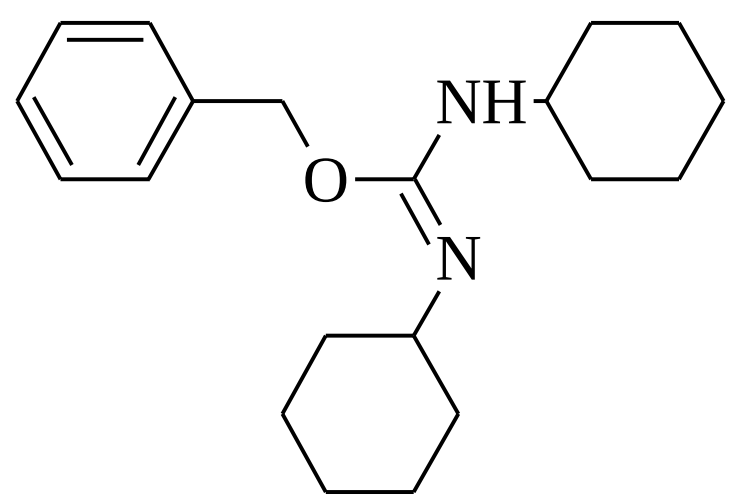

A mixture of N, N'-dicyclohexylcarbodiimide (25.2 g, 120 mmol), benzyl alcohol (13.5 g, 130 mmol), toluene (40 ml), and a bit of copper (I) chloride was stirred at room temperature for 5 days. When the reaction was complete (TLC), toluene evaporated at low pressure, hexane (250 ml) was added, and the mixture passed through an alumina pad. The filtrate was evaporated to dryness. Yield: 34.6 g (88%); $B_p$=130°C/0.0001 Torr.

*Benzyl 2-fluoro-4-hydroxybenzoate* **(3)**

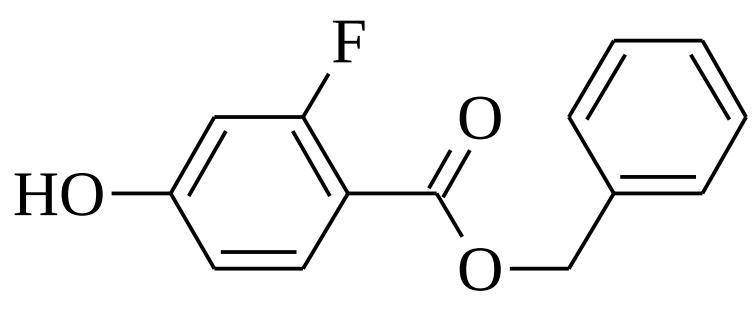

A mixture of 2-fluoro-4-hydroxybenzoic acid (10.6 g, 68 mmol), O-benzyl-N, N'-dicyclohexylisourea (21.4 g, 68 mmol), and dry THF (100 ml) was stirred at room temperature for 6 hours. The precipitate was filtered off, and the filtrate was concentrated at low pressure. The crude product was crystallized from ethanol (100 ml) and hexane=toluene(2:1, 100 ml). Yield: 4.5 g (27%); $M_p$= 90–92°C, GC: 99.8 %; MS(EI) m/z: 246 $[M+H]^+$

*benzyl 2-fluoro-4-[(4-propylbenzoyl)oxy]benzoate* **(4)**

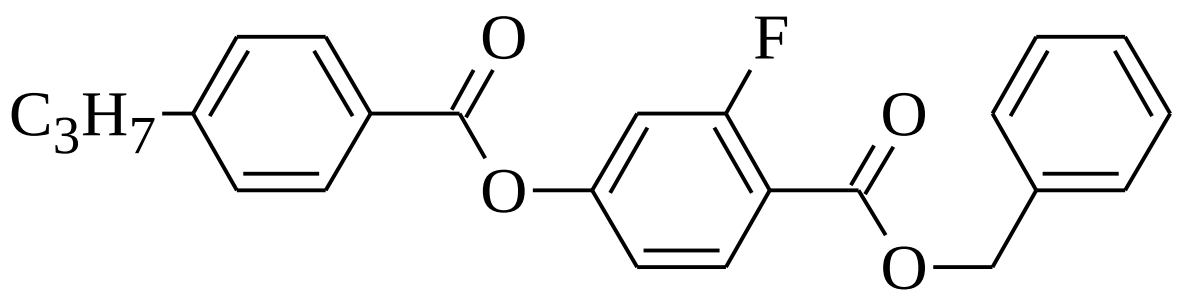

A 250 ml round-bottom flask was charged with benzyl 2-fluoro-4-hydroxybenzoate (2.3 g, 9.3 mmol), pyridine (1.5 g, 18.7 mmol), and dry dichloromethane (35 ml). The 4-propylbenzoyl chloride (1.7 g, 9.3 mmol) was dropped. The reaction mixture was stirred and refluxed until the reaction was complete (TLC). After cooling to room temperature, the reaction mixture was poured into diluted hydrochloric acid and stirred until the precipitate of pyridine hydrochloride was dissolved. Phases were separated. The water phase was extracted with dichloromethane (3 *100 ml). The combined organic fractions were washed with diluted hydrochloric acid (3*100 ml) and water (3*100 ml), dried over $MgSO_4$, and concentrated in a vacuum. The crude product was purified by column chromatography ($SiO_2/CH_2Cl_2$) obtaining colorless oil. Yield: 3.2 g (90%); GC: 98.6 %; MS(EI) m/z: 392 $[M+H]^+$

2-*fluoro-4-[(4-propylbenzoyl)oxy]benzoic acid* **(5)**

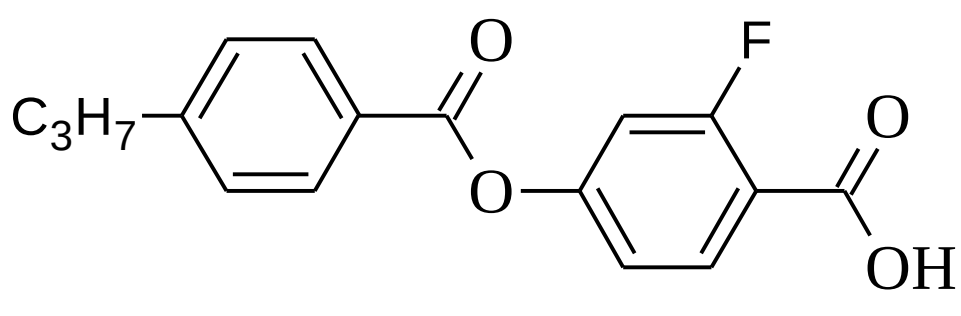

To the solution of benzyl 4-(4-propylbenzoyloxy)-3-fluorobenzoate(3.6 g, 9.2 mmol) in THF (50 ml), 10% palladium on charcoal (0.5 g) was added. The reaction mixture was flushed with hydrogen and connected to a gas burette filled with hydrogen. Stirring was turned on. While hydrogen was consumed, the temperature increased. When the reaction was completed (TLC), the reaction mixture was flushed with nitrogen, and the catalyst was filtered off and rinsed with THF (20 ml). The filtrate was evaporated, and the dry residue was crystallized from ethanol (80 ml). Yield: 1.6 g (56.7%); GC: 99.7%; MS(EI) m/z: 302 $[M+H]^+$

*3,4,5-trifluorophenyl 2-fluoro-4-[(4-propylbenzoyl)oxy]benzoate* **3F**

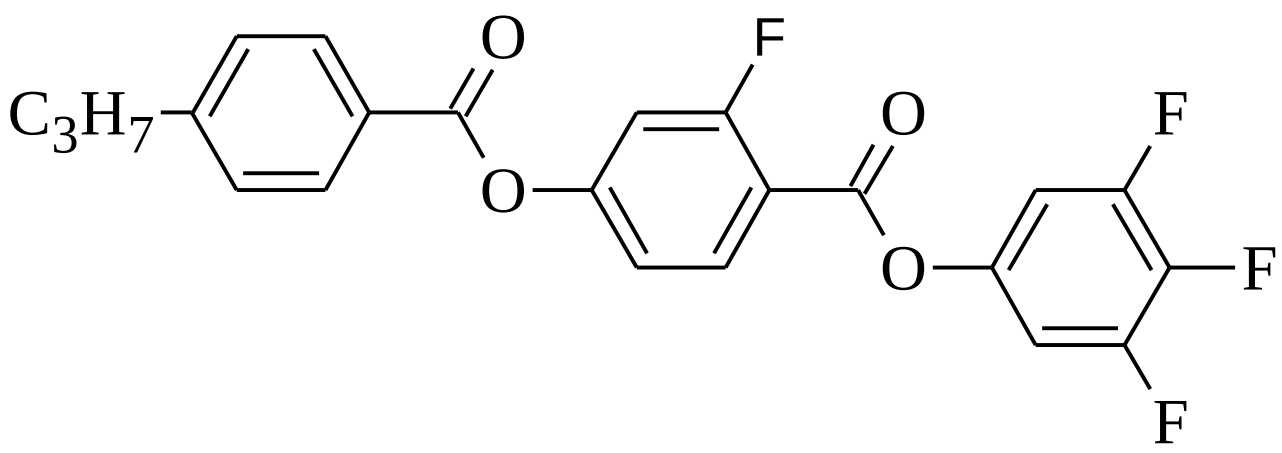

To the suspension of 4-(4-propylbenzoyloxy)-2-fluoro benzoic acid (0.9 g, 2.3 mmol) in dry dichloroethane (50 ml) oxalyl chloride (0.57 g, 4.5 mmol) and one drop of N, N-dimethylformamide was added. The reaction mixture was stirred at room temperature until the evolution of gases stopped. The excess of oxalyl chloride was distilled off with dichloroethane (10 ml). After cooling to room temperature, pyridine (0.7 g, 8.9 mmol), 3,4,5-trifluorophenol (0.44 g, 3.0 mmol), and dichloromethane (10 ml) was added, and the mixture was refluxed for 6 hours. The reaction mixture was cooled to room temperature and poured into diluted hydrochloric acid (50 ml). Phases were separated. The water phase was extracted with dichloromethane (2*50 ml). The combined organic fractions were washed with diluted hydrochloric acid (3*50 ml), water (3*50 ml), and brine (50 ml), dried over MgSO4, and concentrated at low pressure. The crude product was purified by column chromatography ($SiO_2/CH_2Cl_2$) and crystallized from ethanol (50 ml). Yield: 0,7 g (70%); HPLC: 99.8%; MS(EI) m/z: 431 $[M-H]^+$.

$^1$H NMR (500 MHz, $CDCl_3$) δ/ppm: 1.0 (t, *J*=7.3 Hz, 3H, $-CH_3$), 1.7 (td, *J*=15.0, 7.3 Hz, 2H, $-CH_2-$), 2.7 (m, 2H, $-CH_2-$), 7.0 (dd, *J*=7.6, 5.8 Hz 2H, Ar-H), 7.2 (m, 2H, Ar-H), 7.4 (d, *J*=8.2 Hz, 2H, Ar-H), 8.1 (s, 2H, Ar-H), 8.2 (s, 1H, Ar-H).

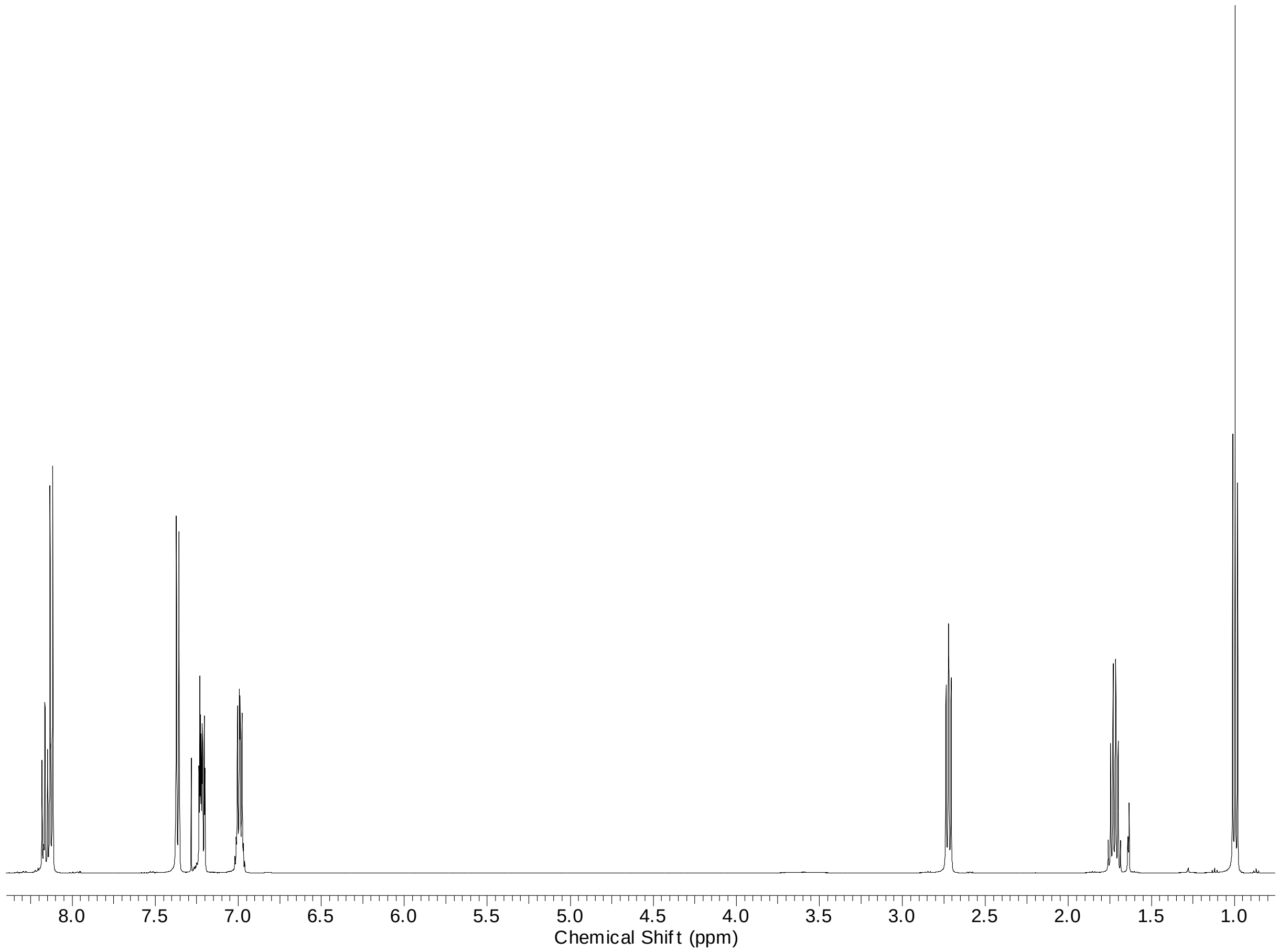

$^{13}C$ NMR (126 MHz, $CDCl_3$) δ/ppm: 13.4 (s, 1C), 23.9 (s, 1C), 37.9 (s, 1C), 106.9 (dd, 2C), 111.1 (s, 2C), 114.1 (d, *J*=9.1 Hz, 1C), 128.7 (s, 3C), 130.1 (s, 3C), 137.0 (t, *J*=15.4 Hz, 1C), 144.8 (td, *J*=11.6, 4.1 Hz, 2C), 149.7 (s, 2C), 149.8 (dd, *J*=10.4, 5.0 Hz, 1C), 161.5 (s, 1C) 163.6 (s, 1C), 163.8 (s, 1 C).

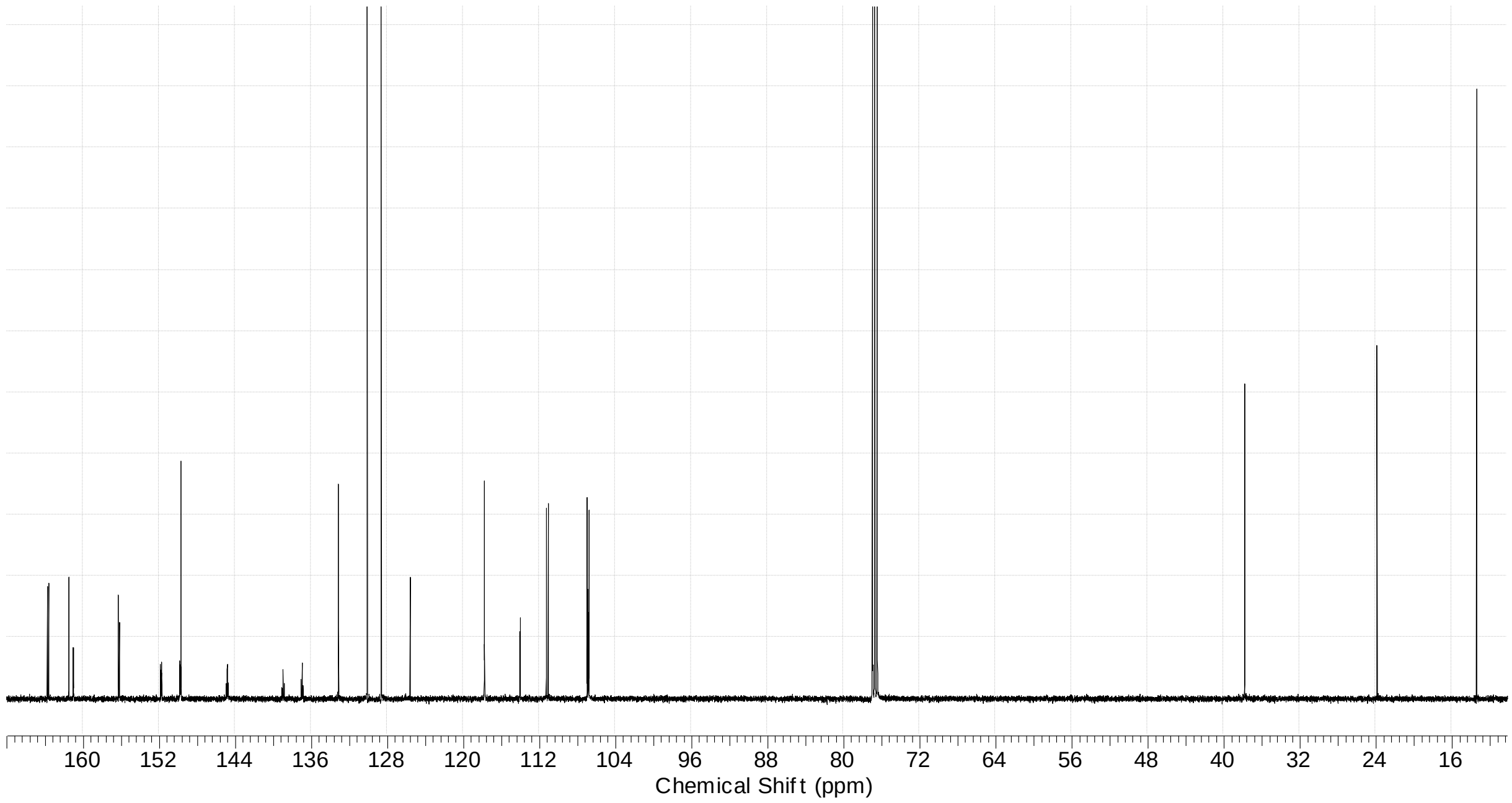

*3,4,5-trifluorophenyl 2-fluoro-4-[(4-butylbenzoyl)oxy]benzoate* **4F**

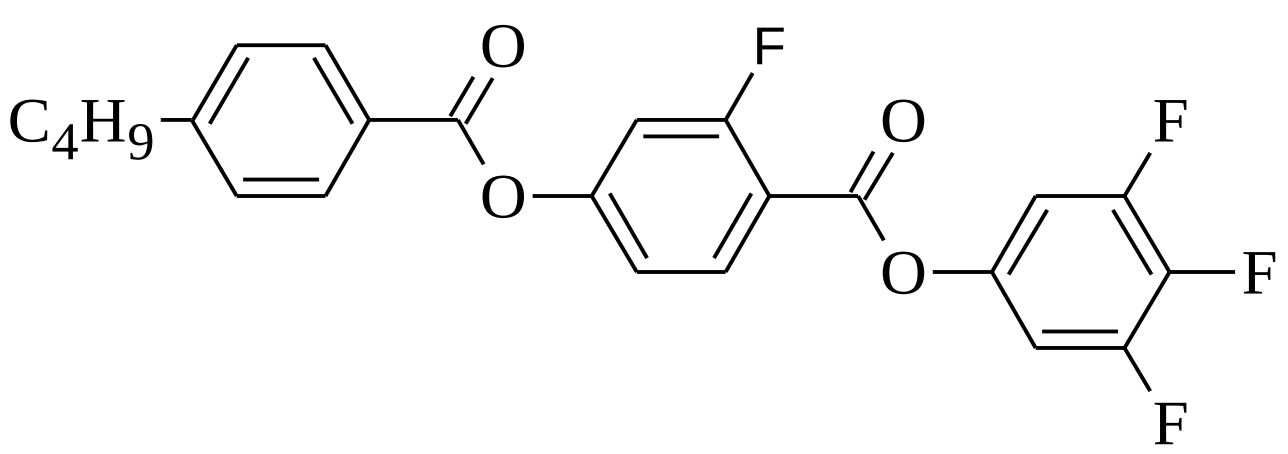

Yield: 0,6 g (59%); HPLC: 99.9%; MS(EI) m/z: 445 $[M-H]^+$.

[1]H NMR (500 MHz, $CDCl_3$) δ/ppm: 0.9 (t, *J*=7.5 Hz, 3H, $-CH_3$), 1.4 (td, *J*=7.8, 7.1 Hz, 2H, $-CH_2-$), 1.6 (m, 2H, $-CH_2-$), 2.7 (m, 2H, $-CH_2-$), 7.0 (dd, *J*=7.6, 5.8 Hz, 2H Ar-H), 7.2 (m, 2H Ar-H), 7.3 (d, *J*=8.2 Hz, 2H Ar-H), 8.1 (d, *J*=8.2 Hz, 2H Ar-H), 8.1 (m, 1H Ar-H).

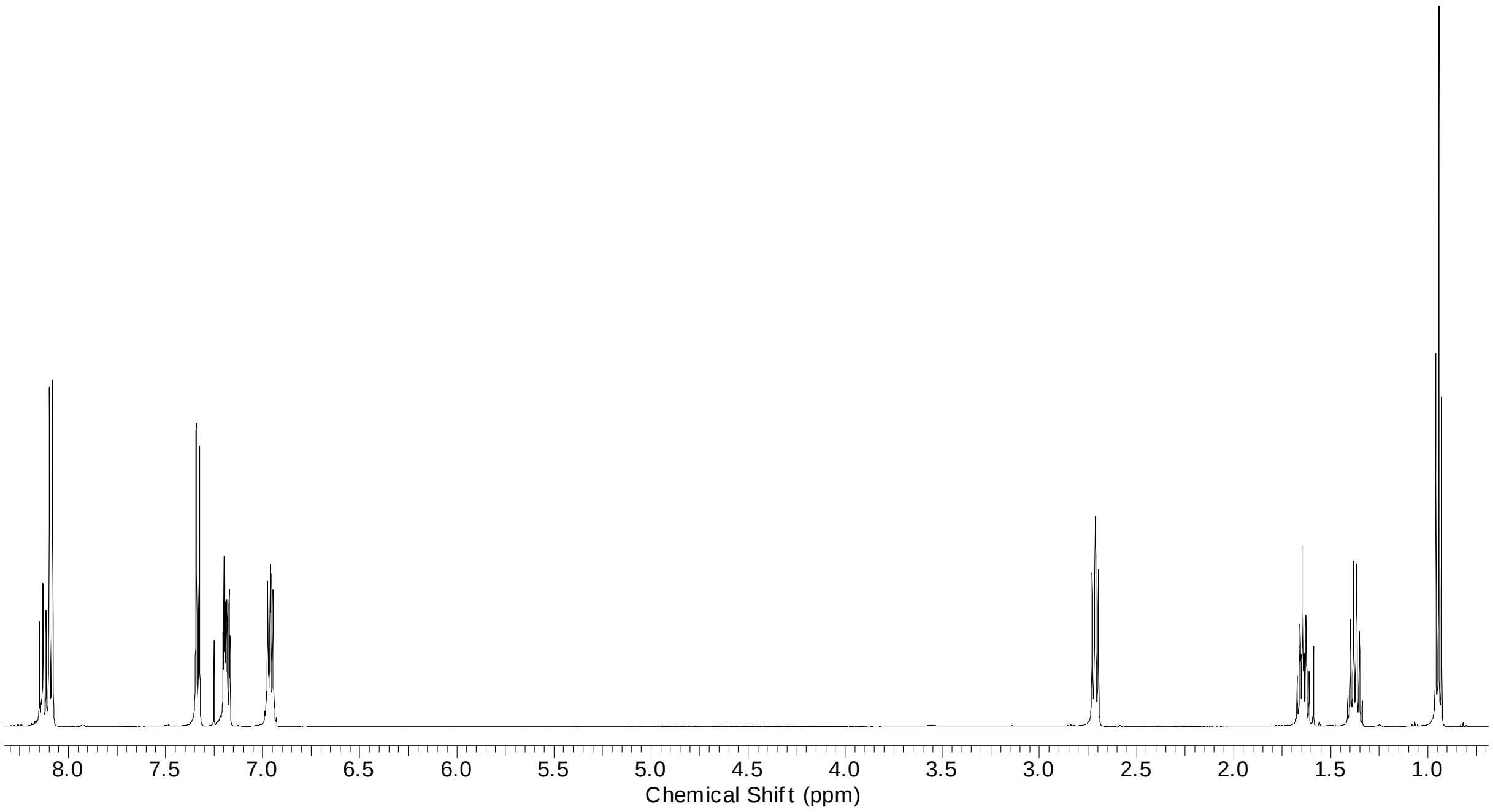

$^{13}C$ NMR (126 MHz, $CDCl_3$) δ/ppm 13.9 (s, 1C), 22.3 (s, 1C), 33.2 (s, 1C), 35.8 (s, 1C), 107.2 (dd, 2C), 111.3 (s, 2C), 114.3 (d, *J*=9.1 Hz, 1C), 128.9 (s, 3C), 130.4 (s, 3C), 137.2 (t, *J*=15.4 Hz, 1C), 145.1 (td, 2C), 150.1 (dd, *J*=10.4, 5.0 Hz, 2C), 161.3 (d, *J*=4.5 Hz, 1C) 161.8 (s, 1C) 163.9 (s, 1C) 164.0 (s, 1C).

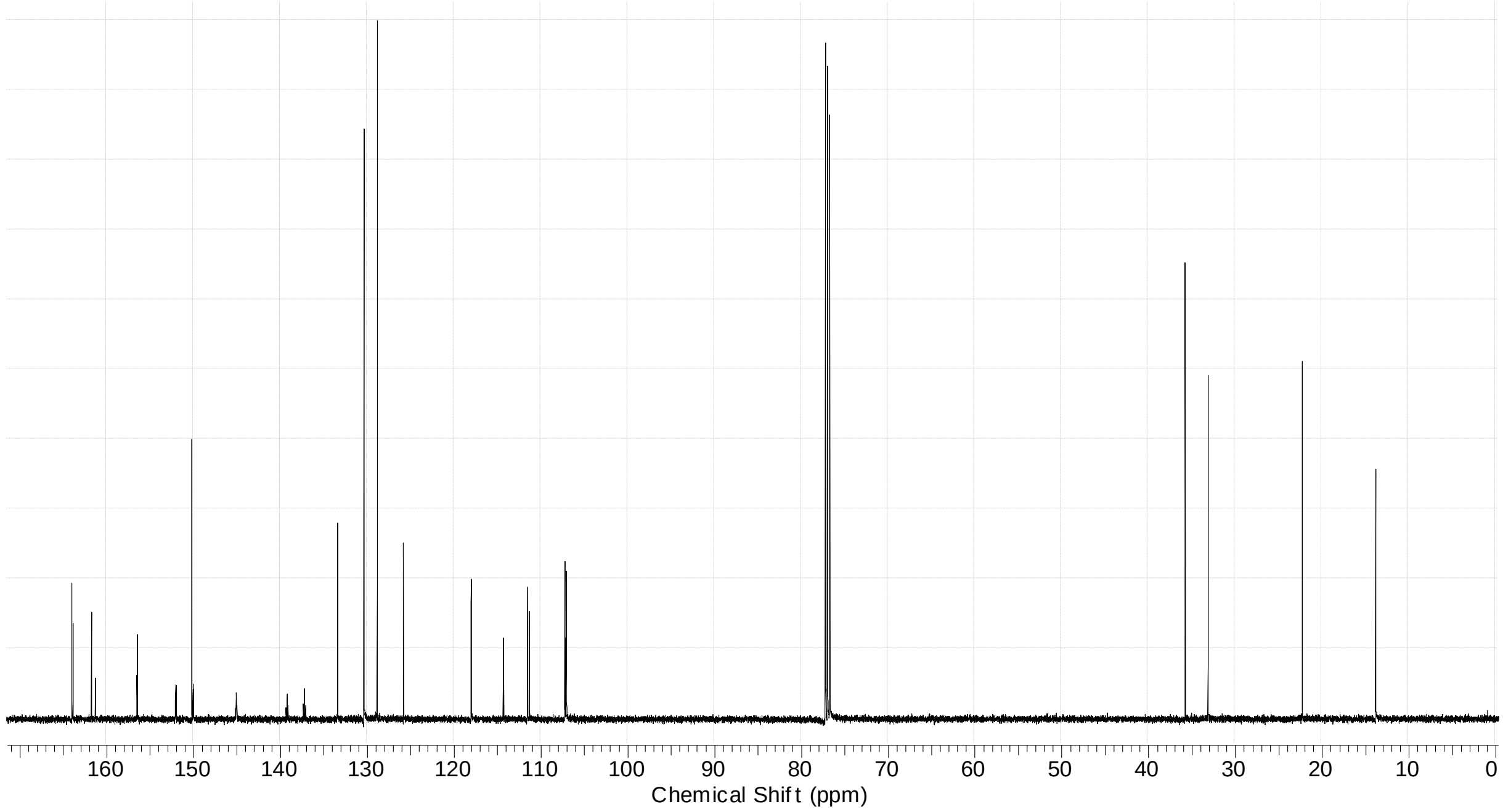

4. Supplemental references